\documentclass[a4paper,twocolumn,11pt]{quantumarticle}
\pdfoutput=1
\usepackage[utf8]{inputenc}
\usepackage[english]{babel}
\usepackage[T1]{fontenc}
\usepackage{amsmath}
\usepackage{hyperref}
\usepackage[numbers,sort&compress]{natbib}
\usepackage{tikz}
\usepackage{lipsum}
\usepackage{url}
\usepackage{epsfig,graphicx,amssymb,color,bm,epstopdf}

\newcommand{\eref}[1]{Eq.~(\ref{#1})}
\newcommand{\esref}[1]{Eqs.~(\ref{#1})}
\newcommand{\fref}[1]{Fig.~\ref{#1}}
\newcommand{\sref}[1]{Sec.~\ref{#1}}

\begin{document}

\title{Photon-phonon quantum cloning in optomechanical system}

\author[1]{Qingxia Mu}
\author[1]{Ting Wang}
\author[2]{Jiong Cheng}
\author[2]{Wen-Zhao Zhang}
\email{zhangwenzhao@nbu.edu.cn}

\affil[1]{Mathematics and Physics Department, North China Electric Power University, Beijing 102206, China}
\affil[2]{School of Physical Science and Technology, Ningbo University, Ningbo 315211, China}

\maketitle

\begin{abstract}
 Quantum cloning is an essential operation in quantum information and quantum computing.
 Similar to the `copy' operation in classical computing, the cloning of flying bits for further processing from the solid-state quantum bits in storage is an operation frequently used in quantum information processing.
 Here we propose a high-fidelity and controllable quantum cloning scheme between solid bits and flying bits. In order to overcome the obstacles from the no-cloning theorem and the weak phonon-photon interaction, we introduce a hybrid optomechanical system that performs both the probabilistic cloning and deterministic cloning closed to the theoretical optimal limit with the help of designed driving pulse in the presence of dissipation.
 In addition, our scheme allows a highly tunable switching between two cloning methods, namely the probabilistic and deterministic cloning, by simply changing the input laser pulse. This provides a promising platform for experimental executability.
\end{abstract}

\section{Introduction}
The quantum no-cloning theorem does not allow for obtaining two identical quantum states by cloning an unknown quantum state \cite{nature.299.802}.
The memory operation is easy to perform in the classical field, but difficult to realize in quantum regime.
However, it is an important premise for quantum information processing and quantum cryptography \cite{RevModPhys.77.1225,Sr.10.11687}.
Many schemes have been proposed to face this limit, which allows us to copy the original quantum information imperfect, probabilistic \cite{PhysRevLett.80.4999,PhysRevLett.106.180404,QIP.14.2271, epjd.72.26,CTP.31.223} or reduced fidelity by adding an auxiliary system  \cite{PhysRevA.54.1844,PhysRevA.62.012302,PhysRevA.65.012304,PhysRevA.67.022317,PhysRevA.99.012324,ScienceAdv.3.e1601915,PhysRevLett.126.060503}.
Various physical systems have been studied to realize quantum cloning, including optical systems \cite{PhysRevA.99.012324,OE.27.32454,PhysRevA.101.010302,PhysRevLett.125.210502,PhysRevLett.126.060503}, atomic systems \cite{QIP.14.2271,AP.373.512}, ion systems \cite{PhysRevA.67.032303,TQE.3.1}, superconducting system \cite{npj.QI.7.44}, and so on.
Among them, optical systems with great potential are relatively active in both theoretical and experimental aspects \cite{PhysRevLett.126.210507,PhysRevLett.126.060503,PhysRevA.101.010302,RevModPhys.77.1225}.
Photons as excellent flying bits that can be used as carriers of quantum information, and thereby many protocols were proposed with this property, such as BB84 \cite{BB84,PhysRevA.85.052310} and B92 \cite{PhysRevLett.68.3121}.
Whereas photons are difficult to store, the solid bits make up for this deficiency well, exhibiting advantages such as easy storage and long decoherence times.
Existing experimental reports can reach even the order of hours in atomic frequency comb \cite{NAT.COMMUN.12.2381}.
Quantum cloning also been implemented in such type systems such as nitrogen vacancy center \cite{nature.469.508} and crystal \cite{ScienceAdv.2.e1600911}.
However, how to combine these two types of systems, that is, to read out a copy of the information in the flying qubit for further processing, while also keeping the original one stored in the solid system remains unchanged, is a challenging task in quantum information processing.

Fortunately, the optomechanical system (OMS) established with the principle of coupling the mechanical oscillator and the optical cavity via the effect of radiation pressure \cite{gigan2006self,schliesser2006radiation}, provides us with a feasible platform to realize quantum interactions between phonon and photon with widely differing frequencies, and thus we can  realize the `copy' operation  mentioned before.
In recent years, the application of OMS in quantum information processing and quantum computing has been widely studied \cite{PhysRevLett.109.013603,PhysRevA.100.052306,PRXQuantum.2.040345}.
Optomechanical sideband cooling stabilizes the phonon to its ground states and thereby shows quantum behavior, which has been well explored both theoretically and experimentally \cite{liu2015coupled,qu2015generating}.
With this quantum behavior, OMS can be used as a platform for quantum information processing, which exhibits Kerr nonlinear effects under certain parameter conditions and further be used as quantum controlled gates or transmission control devices, etc., and even to build fault-tolerant quantum computing \cite{RIP.35.105338,PhysRevA.91.043822}.
In addition, the coupling between photon-phonon can be enhanced by a factor of the square root of the number of photons in the cavity under linearization conditions \cite{bhattacharya2008multiple,xuereb2012strong,newsom2020optimal}, which allows us to observe strong quantum interactions between optical mode and mechanical mode in the presence of environmental noise, including photon-phonon state transfer, entanglement or squeezing \cite{RevModPhys.86.1391}.
Altogether, both the quantum behavior exhibited by the OMS and its application to quantum information processing provide the conditions for us to carry out quantum cloning between solid bits and flying bits (phonon and photon).
In our proposal, as shown in \fref{fig1}, a five-mode optomechanical system is used, which is constructed with two membrane mechanical resonators centered in the Fabry-P\'erot optical cavity and two nano mechanical resonators placed outside the cavity \cite{ludwig2012enhanced,xu2015optical}.
The collective interaction resulting from the coupling of intracavity double-mechanical modes arrays enhances the optomechanical coupling strength \cite{bhattacharya2008multiple,xuereb2012strong,newsom2020optimal}, and multimode circuit optomechanics has been experimentally implemented \cite{massel2012multimode}.
With this hybrid optomechanical system, we designed control pulses to perform linear and nonlinear operations between photon-phonon by controlling the strength of the external drive as well as the coupling of the mechanical oscillators to achieve the `copy' we need.

This paper is organized as follows: In Sec.~\ref{model},
the model and the Hamiltonian of our optomechanical system are proposed.
In Sec.~\ref{effHamiltonian}, the system's effective Hamiltonian is obtained respectively under the weak driving condition and the strong driving condition according to Langevin equations.
The high-fidelity controlled phase flip gates (CPFG) and swap gates for quantum cloning are discussed in this section.
In Sec.~\ref{cloning}, the phonon-photon quantum cloning process is shown in detail.
Finally, Sec.~\ref{conclusion} presents our discussion and conclusion.

\section{Model and physical analysis of nonlinear enhancement}\label{model}
\begin{figure}
  \centering
  \includegraphics[width=7.5cm]{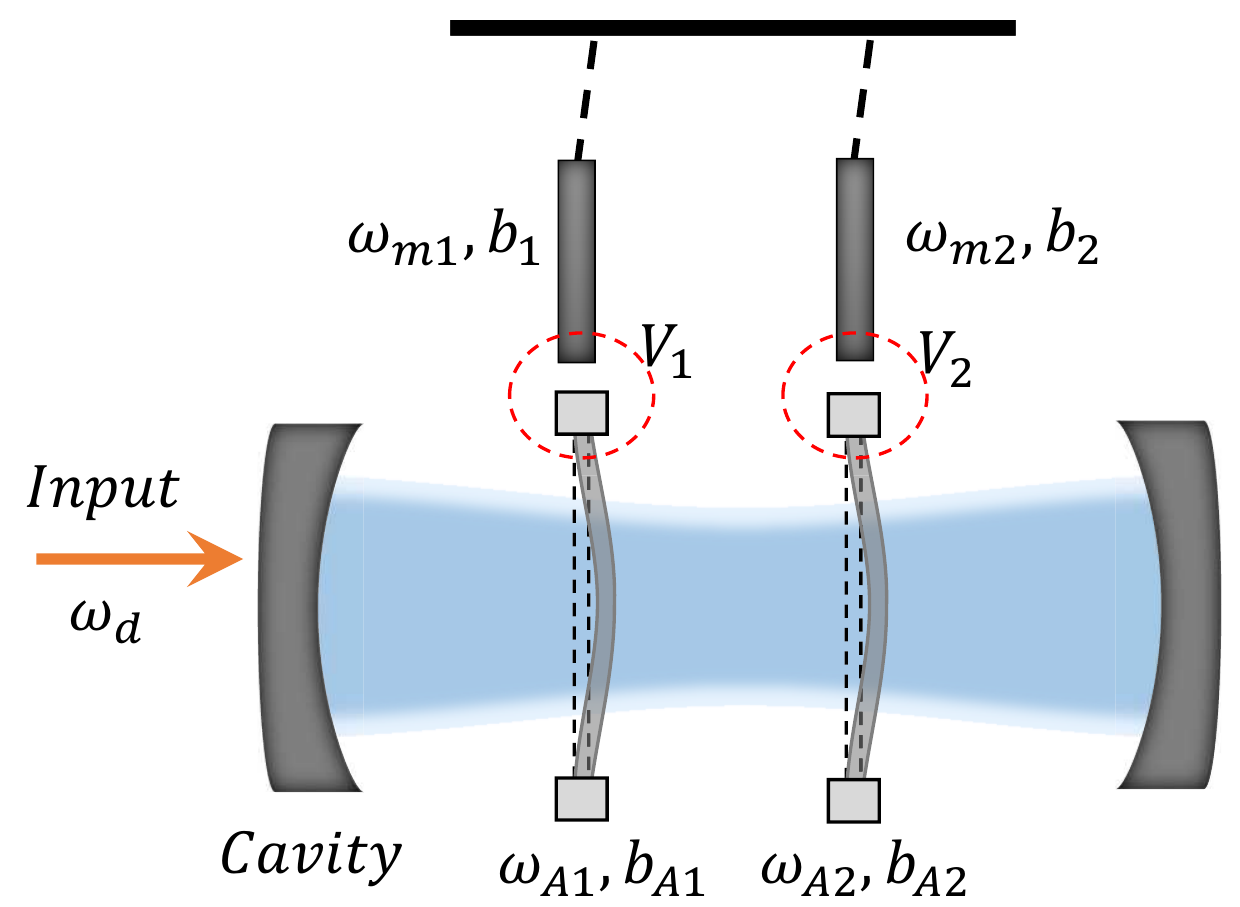}\\
  \caption{The schematic diagram of the five-mode optomechanical system.
  In order to realize the single or bilateral controlled gate, we use the internal mechanical membranes to enhance the nonlinear coupling between the optical and external mechanical modes.}\label{fig1}
\end{figure}

The OMS to perform photon-phonon cloning under consideration is illustrated in \fref{fig1}, where two membranes inside a Fabry-P\'erot cavity are quadratically couples to the cavity field \cite{piergentili2018two,doi:10.1021/acs.nanolett.8b03240,naserbakht2019electromechanics,wei2019controllable}.
To control and adjust the nonlinear coupling between the two mechanical and optical modes separately, each membrane is coupled to an external mechanical oscillator.
The Hamiltonian of the designed system is described as,
\begin{eqnarray}
H &=& \omega_c a^{\dag} a+ \sum_{j=1}^2\Big[ \omega_{Aj}b_{Aj}^{\dag}b_{Aj} -g_j a^{\dag}a(b_{Aj}+b_{Aj}^{\dag})^2 \nonumber \\
&&+ \omega_{mj}b_{j}^{\dag}b_{j}+V_j(b_{Aj}^{\dag}b_{j}+b_{j}^{\dag}b_{Aj})\Big]+H_d,
\end{eqnarray}
where $a$ is the annihilation operator of optical mode with frequency $\omega_c$. $b_{Aj}$ and $b_{j}$ are the annihilation operators for the internal and external mechanical modes with frequency $\omega_{Aj}$ and $\omega_{mj}$, respectively.
The third term in the Hamiltonian represents the quadratic optomechanical coupling between the cavity and the mechanical membranes with strength $g_j$.
The fifth term describes the phonon tunneling coupling between the corresponding oscillators with strength $V_j$.
For simplicity and without loss of generality, we assume that $V_j$ is real.
$H_d$ is the driving term, which is described as $H_d=\epsilon (a^{\dag}e^{-i\omega_d t}+ae^{i\omega_d t})$, where $\epsilon$ denotes the corresponding driving strength and $\omega_d$ is the driving frequency.
With the rotating wave approximation we obtain the Hamiltonian after neglecting the terms $b_{Aj}^{2}$ and $b_{Aj}^{\dag^2}$,
\begin{eqnarray}
H' &=& \omega_c'a^{\dag}a+\sum_{j=1}^2\Big[\omega_{Aj}b_{Aj}^{\dag}b_{Aj}+ \omega_{mj}b_{j}^{\dag}b_{j}\\
&&-2 g_j a^{\dag}a b_{Aj}^{\dag}b_{Aj}+V_j (b_{Aj}^{\dag}b_{j}+b_{Aj}b_{j}^{\dag})\Big]+H_d,\nonumber
\end{eqnarray}
where $\omega_{c}'=\omega_c-g_1-g_2$.
A significant cross-Kerr nonlinearity between photons and phonons ($2g_ja^{\dag}a b_{Aj}^{\dag}b_{Aj}$) is observed in the above Hamiltonian.
However, the corresponding nonlinear parameters are rather weak under the existing experimental conditions \cite{RevModPhys.86.1391}.
This is not sufficient to perform quantum manipulation (high fidelity CPFG) between photons and phonons to accomplish our target task in the presence of dissipation.
Therefore, we introduced additional phonon modes, i.e., $b_j$, in the design for enhancing this cross-Kerr nonlinear effect.

\begin{figure*}
  \centering
  \includegraphics[width=14cm]{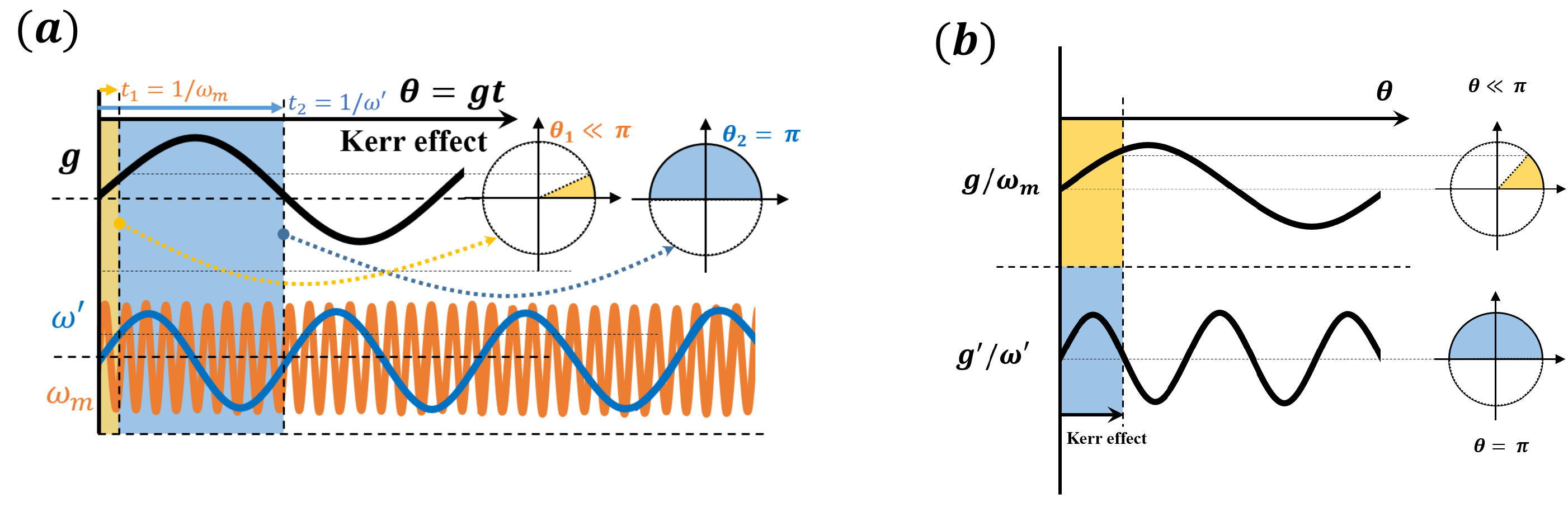}\\
  \caption{Schematic diagram of the Kerr nonlinear effect enhancement by introducing outside modes.}\label{fig2}
\end{figure*}

To show the motivation of our model design, we aim to physically illustrate the mechanism of the introduction of outside modes to enhance the Kerr nonlinear effect.
As reported in Refs.~\cite{PhysRevA.99.063811,PhysRevLett.112.213602}, utilizing the assistant modes to enhance the effective coupling also results in an equivalent enhancement of dissipation.
Thus, if we use the `granularity parameter' \cite{Nphy.4.561} (the ratio of effective coupling to effective dissipation) to estimate the strength of the coupling, the use of outside modes cannot essentially enhance the Kerr nonlinearity parameters.
But to achieve CPFG, we need that the Kerr nonlinearity induced interaction could produce an additional $\pi$ phase in the dynamics \cite{OE.23.7786}.
In the absence of outside modes, the original Kerr coefficient $g_j$ is much smaller than the eigenfrequency $\omega_{Aj}$ and the corresponding Kerr phase can be basically neglected, so it is difficult to achieve a large phase accumulation in a few cycles of the eigenevolution and thus resist the effect of dissipation.
This can also be seen in the upper part of \fref{fig2}~(a) or (b), i.e., the part marked in yellow.
Thus, for this time-accumulated phase, we cannot consider the value of the Kerr nonlinear coefficient independently, which needs to be compared with the eigenfrequency of the mechanical modes (this can also be seen in the expression for the phase factor in \eref{psif} in Sec.~3).
When we introduce the outside modes, the BS-type oscillator coupling $V_j$ will cause a mode splitting, i.e., the so-called quasi-modes \cite{PhysRevLett.109.063601}.
In this case, the coupling resonator has a high and a low effective frequency.
Since the quasi-modes at low frequencies are chosen ($\omega_j'$), the eigenfrequencies of the dressed resonators can be comparable to the Kerr coefficients.
So that the phase accumulated by the Kerr effect cannot be neglected in the effective dynamics, and it becomes easy to achieve large phase accumulation in a few eigencycles.
\fref{fig2}~(a) or (b) are plotted to provide two different ways to illustrate this physical process, respectively.
\fref{fig2}~(a) represents a schematic of the phase accumulation of the Kerr coefficients, i.e., the original frequency and the low-frequency quasi-mode over time.
The accumulation of the Kerr phase can be simply expressed as the product of the Kerr coefficient and time.
When $g_j \ll \omega_{Aj}$, the accumulated phase in one cycle is much less than $\pi$, while when the frequency is reduced using the quasi-mode $\omega_j'$, the accumulated phase in one cycle can reach $\pi$.
\fref{fig2}~(b) illustrates it from the viewpoint of relative frequency, and the Kerr phase can be expressed as the product of the relative frequency and the number of eigenperiods, which is essentially the same as in \fref{fig2}~(a).
Thus, when there is no external mode, $g_j/\omega_{Aj}$ is very small and only a small phase can be accumulated in the same periods.
While, when the frequency is reduced by using the outside modes, $g_j'/\omega_j'$ is large enough and the $\pi$ phase can be achieved in the same periods, and then CPFG can be realized.
Thus, for effective dynamics, only a few cycles of eigenevolution are needed to reach the $\pi$ phase.
This also decreases the time accumulation caused by effective dissipation.
We also confirm this conclusion by numerical simulations in \fref{figa1} in the Appendix.
Next, we will specifically show this effective physical process through the mathematical solution of the Hamiltonian.

\section{Effective Hamiltonian and quantum logic gate realization}\label{effHamiltonian}
To implement quantum cloning of phonons in storage, we need to prepare quantum logic gates other than general quantum gates to construct quantum circuits, in which the controlled phase flip gate with $\pi$ phase occupies a key position (In particular, the significant frequency gaps between mechanical-optical modes in our expected target task make it more difficult to achieve high-fidelity quantum logic gates between phonons and photon).
According to common quantum cloning protocols \cite{PhysRevLett.80.4999,PhysRevLett.106.180404,QIP.14.2271, epjd.72.26,CTP.31.223,PhysRevA.54.1844,PhysRevA.62.012302,PhysRevA.65.012304,PhysRevA.67.022317,PhysRevA.99.012324,ScienceAdv.3.e1601915,PhysRevLett.126.060503}, three kinds of CPFG need to be implemented based on the model given in \fref{fig1}, i.e., photon-phonon CPFG, photon-phonons bilateral CPFG and phonon-phonon CPFG.

A very effective way to construct these CPFGs are using the quantum Kerr effect, which has been extensively studied in optomechanical systems, where the system can exhibit equivalent Kerr nonlinear effects under appropriate parameters \cite{OE.25.10779,asjad2015quantum,RIP.35.105338,PhysRevA.91.043822}.
Using this effect, by choosing the input and output time, we can achieve the photon-phonon and phonon-phonon quantum CPFG we need.
Now, we will show in detail how CPFGs are implemented in our model.
In our system, we use modes $b_{Aj}$ as auxiliary modes to enhance the cross-Kerr nonlinearity.
$a$ and $b_j$ are used as the modes that we need to be manipulated in quantum cloning.
The corresponding initial arbitrary state of quantum cloning in our system is expressed as
\begin{eqnarray}
|\Psi\rangle _{a,b_1,b_2} = \sum_{j=1}^8 \alpha_j|c_j\rangle,
\end{eqnarray}
As shown in Table~\ref{table1}, $|c_j\rangle$ express the direct product points of $a,b_{1},b_{2}$ arranged in positive order from $|0\rangle$ to $|1\rangle$.
$|c_1\rangle$ denotes $|000\rangle_{a,b_1,b_2}$, which means no photon in the cavity and no phonon in the mechanical oscillators.
$|c_8\rangle$ denotes $|111\rangle_{a,b_1,b_2}$, which means one photon in the cavity and one phonon in each mechanical oscillator.
The corresponding superposition coefficients satisfy the condition $\sum\limits_{j=1}^8|\alpha_{j}|^{2}=1$.
Three kinds of CPFG need to be implemented are also displayed in Table~\ref{table1}.
\begin{table*}[h!t]
    \centering
    \begin{tabular}{|c|c|c|c|c|}
    \hline
    Marker & Basis & $\text{CPFG}_{a,b_1}$ & $\text{CPFG}_{a,b_2}$ & $\text{CPFG}_{a,b_1b_2} $ \\
    \hline
    $|c_1 \rangle$  & $|000\rangle_{a,b_1,b_2}$ & $-$ & $-$ & $-$\\
    \hline
    $|c_2 \rangle$  & $|001\rangle_{a,b_1,b_2}$ & $-$ & $-$ & $-$ \\
    \hline
    $|c_3 \rangle$  & $|010\rangle_{a,b_1,b_2}$ & $-$ & $-$ & $-$ \\
    \hline
    $|c_4 \rangle$  & $|011\rangle_{a,b_1,b_2}$ & $-$ & $-$ & $-$ \\
    \hline
    $|c_5 \rangle$  & $|100\rangle_{a,b_1,b_2}$ & $-$ & $-$ & $-$ \\
    \hline
    $|c_6 \rangle$  & $|101\rangle_{a,b_1,b_2}$ & $-$ & $e^{-i \pi}$ & $-$ \\
    \hline
    $|c_7 \rangle$  & $|110\rangle_{a,b_1,b_2}$ & $e^{-i \pi}$ & $-$ & $-$ \\
    \hline
    $|c_8 \rangle$  & $|111\rangle_{a,b_1,b_2}$ & $e^{-i \pi}$ & $e^{-i \pi}$ & $e^{-i \pi}$ \\
    \hline
    \end{tabular}
    \caption{The markers of the bases vectors and the transformation of the bases vectors by different $\text{CPFG}_{x,y}$.
    Where `$-$' denotes the base have `no change' and `$e^{-i \pi}$' denotes the base attached with the phase of $\pi$.
    The subscripts $x$ and $y$ denote the control and controlled bits of the CPFG, respectively.}\label{table1}
\end{table*}

\subsection{Nonlinearity enhancement under weak driving condition}\label{sec3a}
To utilize the nonlinear effects of the system, we cannot use the so-called linearized approximation \cite{RevModPhys.86.1391}, but instead a weak driving condition is needed.
Under the weak driving approximation, it is safe to ignore the effect of the driving strength in the dynamic.
In the rotating frame of the driving field $\omega_d$, the nonlinear Langevin equations of the system operators can be derived as
\begin{eqnarray}\label{eqls}
\dot{a} &=& -(i \Delta_c'+\frac{\kappa}{2})a+2i \sum_j g_j a b_{Aj}^{\dag}b_{Aj}+\sqrt{\kappa} a_{in},\nonumber\\
\dot{b}_{Aj} &=& -(i\omega_{Aj}+\frac{\gamma_{Aj}}{2}-f_N)b_{Aj}-iV_jb_{j}+\sqrt{\gamma_{Aj}}b_{in}^{Aj},\nonumber\\
\dot{b}_{j} &=& -(i \omega_{mj}+\frac{\gamma_{j}}{2}) b_{j}-iV_jb_{Aj}+\sqrt{\gamma_{j}}b_{in}^{j},
\end{eqnarray}
where $\Delta_c'=\omega_c'-\omega_d$ is the modified detuning of the cavity with a driving frequency $\omega_d$.
$f_N=2ig_ja^{\dag}a$, in which $2g_ja^{\dag}a$ denotes the modified frequency from the optomechanical coupling.
$\kappa$, $\gamma_{Aj}$ and $\gamma_{j}$ are the decay rates of the cavity mode, internal and external mechanical modes, respectively.
$a_{in}$, $b_{in}^{Aj}$ and $b_{in}^{j}$ are the corresponding input noise operators that satisfy $\langle a_{in}(t)a_{in}^{\dag}(t')\rangle=\delta(t-t')$, $\langle b_{in}^{Aj}(t)b_{in}^{Aj\dag}(t')\rangle=(n_{th}^{Aj}+1)\delta(t-t')$ and $\langle b_{in}^{j}(t)b_{in}^{\dag}(t')\rangle=(n_{th}^{j}+1)\delta(t-t')$ under the Bonn-Markov approximation.
$n_{th}=[\exp(\hbar \Omega/k_B T)-1]^{-1}$ is the equivalent thermal occupation and $\Omega$ denotes the eigenfrequency of the thermal phonon.
The formal integral of the operators $b_{Aj}(t)$ and $b_{j}(t)$ according to \eref{eqls} are expressed as
\begin{widetext}
\begin{subequations}
\begin{eqnarray}\label{eqbAt}
b_{Aj}(t) &=& b_{Aj}(0)e^{-(i\omega_{Aj}+\frac{\gamma_{Aj}}{2}-f_N)t}+\int_{0}^{t}d\tau e^{-(i\omega_{Aj}+\frac{\gamma_{Aj}}{2}-f_N)(t-\tau)} (-iV_{j}b_{j}+\sqrt{\gamma_{Aj}}b_{in}^{Aj}),\\
b_{j}(t) &=& b_{j}(0)e^{-(i\omega_{mj}+\frac{\gamma_{j}}{2})t}+\int_{0}^{t}d\tau e^{-(i\omega_{mj}+\frac{\gamma_{j}}{2})(t-\tau)}\left(-iV_j b_{Aj}+\sqrt{\gamma_{j}}b_{in}^{j}\right).\label{eqbt}
\end{eqnarray}
\end{subequations}
\end{widetext}
According to our analysis in Sec.~\ref{model}, we need to dress $b_{j}$ with $b_{Aj}$ and chosen low-frequency eigenmodes.
To eliminate $b_{Aj}$ and thus achieve this goal, we need to solve it approximately.
Here, we use the zero-order approximation of $b_j$ taken to solve $b_{Aj}$ under the first-order approximation, and then substitute back to \esref{eqls} to obtain the effective dynamic equations without $b_{Aj}$.
Similar methods have been used in Refs.~\cite{OE.25.10779,PhysRevLett.112.213602}.
Under the condition $\omega_{mj}\gg \{V_j,\gamma_{j}\}$, one can obtain the zero-order approximate solution of $b_{j}(t)$,
\begin{equation}
b_{j}^{(0)}(t)\approx b_{j}(0)e^{-(i\omega_{mj}+\gamma_{j}/2)t}+B_{in}^{j}(t),
\end{equation}
where $B_{in}^{j}(t)=\int_{0}^{t}d\tau e^{-(i\omega_{mj}+\frac{\gamma_{j}}{2})(t-\tau)}\sqrt{\gamma_{j}}b_{in}^{j}\left(\tau\right)$.
Substituting the expression of $b_{j}^{(0)}(t)$ into $b_{Aj}(t)$, the first-order solution can be obtained (see Appendix~\ref{ASPT} for details),
\begin{eqnarray}
b_{Aj}(t)&\approx &b_{Aj}(0)e^{-(i\omega_{Aj}+\gamma_{Aj}/2-f_N)t}+B_{in}^{Aj}(t)\nonumber \\ &&+\frac{-iV_j b_{j}(t)}{i(\omega_{Aj}-\omega_{mj})+\frac{\gamma_{Aj}-\gamma_{j}}{2}-f_N},
\end{eqnarray}
where $B_{in}^{Aj}(t)=[\sqrt{\gamma_{Aj}}b_{in}^{Aj}-i V B_{in}^{j}(t)]/(i\omega_{Aj}+\gamma_{Aj}/2-f_N)$ indicates the input noise after modification.
When $\gamma_{Aj}\gg\gamma_{j}$, the term containing $e^{-\gamma_{Aj}/2}$ is a fast decaying term and can be
neglected.
Under conditions of weak coupling and weak driving, we have $f_N \ll A$.
The above formula can be approximated as
\begin{eqnarray}
b_{Aj}(t) & \approx & \frac{-iV_j b_{j}(t)}{A} \left\{ 1+\frac{f_N}{A}+O[\frac{f_N}{A}]^2 \right\}+B_{in}^{Aj}(t),\nonumber \\ \label{eqbj}
\end{eqnarray}
where $A=i(\omega_{Aj}-\omega_{mj})+(\gamma_{Aj}-\gamma_{j})/2$.
Substituting \eref{eqbj} into \esref{eqls}, we can obtain the dynamic equations under first-order approximation,
\begin{subequations}\label{eqeffl}
\begin{eqnarray}
\dot{a}&=& -(i\Delta_{c}'+\frac{\kappa}{2})a+i\sum_jg_{j}'a b_{j}^{\dagger}b_{j}+\sqrt{\kappa}a_{in}, \\
\dot{b}_{j}&=&-(i\omega_{j}'+\frac{\gamma_{j}'}{2})b_{j}+i g_{j}'a^{\dagger}a b_{j}+\sqrt{\gamma_{j}'} \mathcal{B}_{in}^{j},
\end{eqnarray}
\end{subequations}
where $\mathcal{B}_{in}^{j}=b_{in}^{j}-iV_j B_{in}^{Aj}/\sqrt{\gamma_j}$ denotes the effective input noise operator, the effective parameters are
\begin{subequations}\label{paras}
\begin{eqnarray}
    \omega_{j}'&=& \omega_{mj}-\delta_j(\omega_{Aj}-\omega_{mj}),\\
    g_{j}' &=& 2 \delta_j g_j,\\
    \gamma_j' &=& \gamma_{j}+\delta_j (\gamma_{Aj}-\gamma_{j}),\\
    \delta_j &=& |V_j|^2/|A|^2.
\end{eqnarray}
\end{subequations}
Thus, the corresponding effective Hamiltonian is written as,
\begin{equation}\label{heff}
H_{eff}=\Delta_{c}' a^{\dag}a+\sum_{j=1}^2 \left(\omega_j' b_{j}^{\dag}b_{j}-g_{j}' a^{\dag}a b_{j}^{\dag}b_{j}\right).
\end{equation}

\begin{figure}
  \centering
  \includegraphics[width=8.5cm]{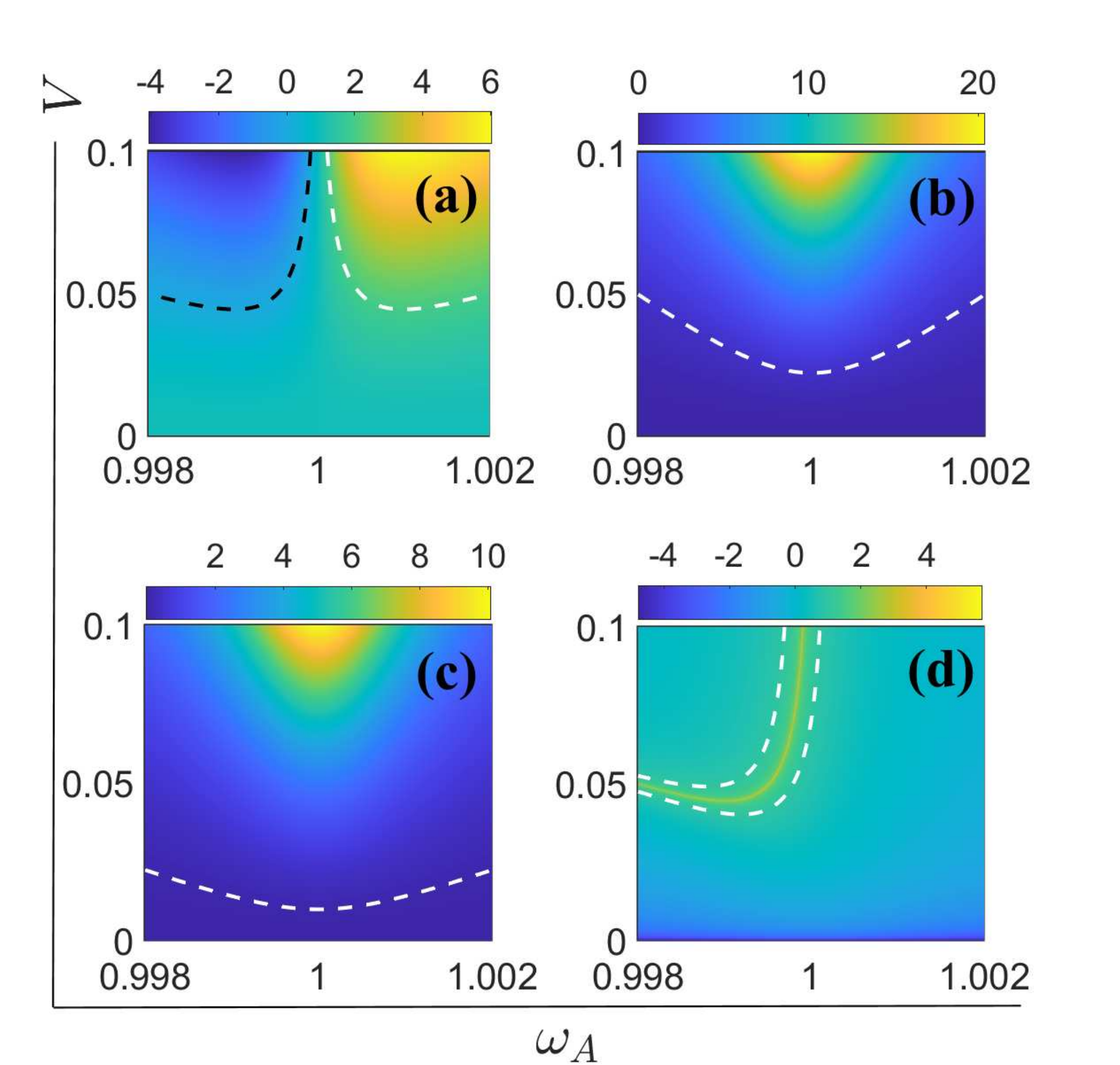}\\
  \caption{The effective parameters as a function of $V$ and $\omega_{A}$ (density plotted in units of $\omega_m$).
  (a) The density plot of $\omega_{j}'$.
  (b) The density plot of $g_{j}'$.
  (c) The density plot of $\gamma_{j}'$.
  (d) The density plot of $\log(g_{j}'/\omega_{j}')$.
  The parameters are $\gamma/\omega_m=10^{-5}$, $\gamma_A/\omega_m=10^{-3}$, $g/\omega_m=10^{-3}$ and $\omega_m=2\times 10^6 \text{Hz}$.
  }\label{fig3}
\end{figure}

For simplify, we set $\omega_{mj}=\omega_m$, $\omega_{Aj}=\omega_{A}$, $\gamma_{j}=\gamma$, $\gamma_{Aj}=\gamma_{A}$, $V_j=V$ and $g_j=g$.
In our scheme, the importance is the phase accumulation caused by the nonlinear interaction.
Thus, according to the analysis in \fref{fig2}, the relative values of nonlinear parameters are more important to us, i.e., $g_{j}'/\omega_{j}'$, $g_{j}'/\kappa$ and $g_{j}'/\gamma_j'$.
The relevant effective parameters and the relationship between them and original parameters are shown in \fref{fig3}, where $g_{j}'/\omega_{j}'$ directly reflects the proportion of nonlinear effects in the effective Hamiltonian.
If this ratio is larger, the dynamics dominated by this Hamiltonian can realize the accumulation of nonlinear effects faster, that is, phase accumulation.
According to \fref{fig3}~(a) and (b), we can see that the effective frequency can even be reduced to $0$, and the effective coupling can be $20$ times of the intrinsic frequency of the oscillator, which means, we can easily make the nonlinear effect dominate the dynamics (see \fref{fig3}~(d), the area enclosed by the white dotted line).
Besides, the proportional relationship between $g_{j}'$ and $\{ \kappa, \gamma_j'\}$ directly reflects the robustness of dynamics to dissipation.
The larger their ratio, the better the dissipation can be ignored.

For the dissipation of the optical cavity $\kappa$, it is worth noting that the whole effective process, that is, the elimination of the optomechanical oscillator, has no direct effect on the optical mode.
According to \esref{eqeffl}, the value of $\kappa$ is not changed during the effective process, while the commonly observed dissipation in the experiment is less than the eigenfrequency of the oscillator, e.g. $\kappa< \omega_m$ \cite{RevModPhys.86.1391}.
Since the effective coupling can reach $20$ times of the intrinsic frequency of the oscillator ($\omega_m$) in our scheme, the condition $g_{j}'=2\delta_j g_j>\kappa$ can be obtained by selecting appropriate parameters.
Thus, the effective Kerr nonlinearity can resist the effect of optical dissipation within a certain period in the dynamic of the system.

However, for the effective mechanical dissipation $\gamma_j'$, is relatively difficult to eliminate.
This result can be obtained from the  expression of effective dissipation in \esref{paras}.
Like the effective coupling, the effective dissipation is also proportional to the proportional coefficient $\delta_j'$.
By comparing \fref{fig3}~(b) and (c), it can be seen that there is a similar parameter dependence between effective coupling and effective dissipation.
The effective dissipation grows as faster as the effective coupling.
It is not surprising, because when we use the strong dissipation auxiliary oscillator to enhance the nonlinear coupling, the dissipation of the auxiliary oscillator can not be ignored, which will inevitably make the dissipation be equally enhanced in the effective process.
Under the condition of eliminating the mode $b_{Aj}$, we have $\gamma\ll  \{ \gamma_{A}, \delta_j \}$.
Then we can obtain that $g_{j}'/\gamma_j' \approx 2g_j/\gamma_{A}$=2 (using the parameters in \fref{fig3}).
In this way, a strong enough effective Kerr nonlinear effect can be obtained to realize the high-fidelity CPFG.
When $|\Psi\rangle_{a,b_1,b_2}$ passes through the photon-phonon $\text{CPFG}_{a,b_1}$, the output state is expressed as (see Table~\ref{table1})
\begin{align}
\left|\Psi'\right\rangle _{a,b_1,b_2}	\!=\!\sum_{j=1}^8\alpha_j|c_j\rangle-2\alpha_{7}\left|c_7\right\rangle-2\alpha_{8}\left|c_8\right\rangle.
\end{align}
According to the $H_{eff}$, after ignoring the cavity's decay and the mechanical damping, the final state can be described as
\begin{eqnarray}\label{psif}
|\Psi_f\rangle_{a,b_1,b_2}=\sum_{j=1}^8\alpha_je^{-i\mu_jt}|c_j\rangle,
\end{eqnarray}
where $\mu_{1}=0$, $\mu_{2}=\omega_{2}'$, $\mu_{3}=\omega_{1}'$, $\mu_{4}=\omega_{1}'+\omega_{2}'$, $\mu_{5}=\Delta_{c}'$, $\mu_{6}=\Delta_{c}'+\omega_{2}'-g_{2}'$, $\mu_{7}=\Delta_{c}'+\omega_{1}'-g_{1}'$, $\mu_{8}=\Delta_{c}'+\sum_j (\omega_{j}'-g_{j}')$.

\begin{figure}
  \centering
  \includegraphics[width=8cm]{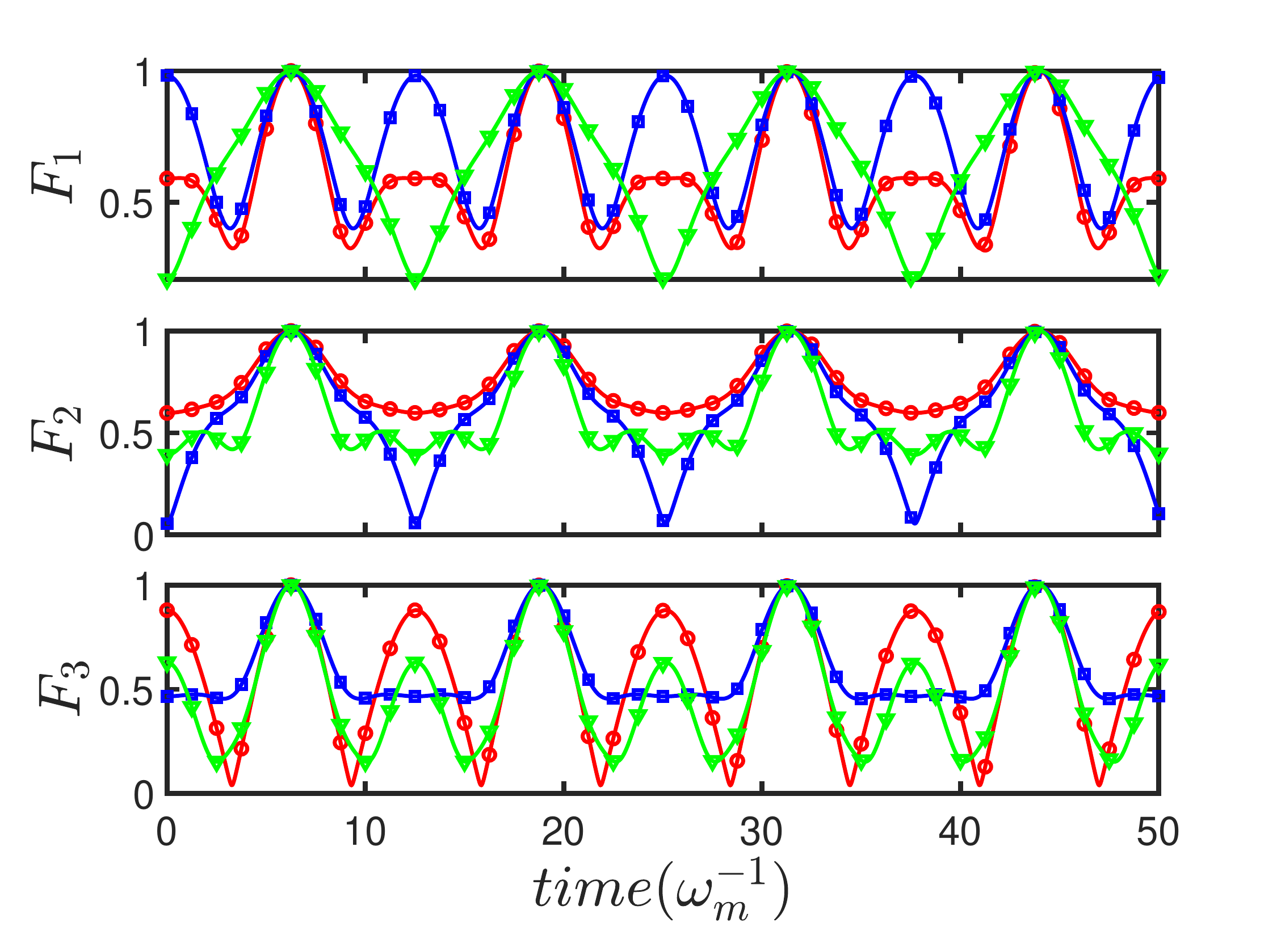}\\
  \caption{The fidelity of controlled phase gate.
  $F_j$ is the fidelity of the CPFG expressed in \esref{eqf1}-\eqref{eqf3}.
  Curves of different colors represent different initial states generated randomly.
  The corresponding color markers indicate the results of the numerical simulation.
  Other parameters are $g/\omega_m=10^{-3}$, $V/\omega_m=0.046$, $\omega_A/\omega_m=0.998$,
  and $\Delta_c'=2 \omega_m$.
  For $F_1$, we choose $g_1=g$ and $g_2=0$.
  For $F_2$, we choose $g_1=0$ and $g_2=g$.
  For $F_3$, we choose $g_1=g$ and $g_2=g$.}\label{fig4}
\end{figure}

We usually use fidelity as a merit factor, which can be defined as $F_{1}=\left|\left\langle \Psi_{f}|\Psi'\right\rangle_{a,b_1,b_2} \right|$.
According to the expression of $|\Psi_f\rangle_{a,b_1,b_2}$, we have
\begin{equation}\label{eqf1}
F_1=\sum_{j=1}^8|\alpha_{j}|^{2}e^{-i\mu_{j}t}-2|\alpha_{7}|^{2}e^{-i\mu_{7}t}-2|\alpha_{8}|^{2}e^{-i\mu_{8}t},
\end{equation}
and when $F_{1}=1$, we have $\mu_{2}t=\mu_{3}t=\mu_{4}t=\mu_{5}t=\mu_{6}t=2n\pi$, and $\mu_{7}t=\mu_{8}t=(2n+1)\pi$.
Where $n$ is an integer.
Similarly, if we want to control phonon $b_2$ with photon, we just need to change the target state to $|\Psi'\rangle_{a,b_1,b_2}=\sum_{j=1}^8\alpha_j|c_j\rangle-2\alpha_{6}|c_7\rangle-2\alpha_{8}|c_8\rangle$
and repeat the above process to obtain the corresponding fidelity $F_2$ between the arbitrary initial state after passing through our system and the target state,

\begin{equation}\label{eqf2}
F_2=\sum_{j=1}^8|\alpha_{j}|^{2}e^{-i\mu_{j}t}-2|\alpha_{6}|^{2}e^{-i\mu_{7}t}-2|\alpha_{8}|^{2}e^{-i\mu_{8}t}.
\end{equation}
Again, replacing the target state with $|\Psi'\rangle_{a,b_1,b_2}=\sum_{j=1}^7\alpha_j|c_j\rangle-\alpha_{8}|c_8\rangle$, we can derive the fidelity of the bilateral control gate in which photon simultaneously control phonon $b_1$ and $b_2$ for phase flip ($\text{CPFG}_{a,b_1b_2}$) as,
\begin{equation}\label{eqf3}
F_3=\sum_{j=1}^7|\alpha_{j}|^{2}e^{-i\mu_{j}t}-|\alpha_{8}|^{2}e^{-i\mu_{8}t}.
\end{equation}
Similarly, the corresponding type of CPFG is realized when the appropriate dynamic time is selected and makes $F_3=1$. Neglecting dissipation, the fidelity of photon-phonon CPFG and photon-phonons bilateral CPFG driven by effective Hamiltonian as a function of time is displayed in \fref{fig4}.
It is observed that the CPFG is achieved periodically at a specific moment as long as the appropriate parameters are adjusted.
The parameters are chosen here based on the analytical solution of \eref{eqf1}.
We set $g_j'/\omega_j'= 0.5$, corresponding original parameters are obtained from \fref{fig3}, i.e., $g/\omega_m=10^{-3}$, $V/\omega_m=0.046$ and $\omega_A/\omega_m=0.998$.
This realization time is independent of the selection of the initial state.
The analytical and numerical comparisons are also displayed in \fref{fig4} by the corresponding color markers.

\begin{figure}
  \centering
  \includegraphics[width=8cm]{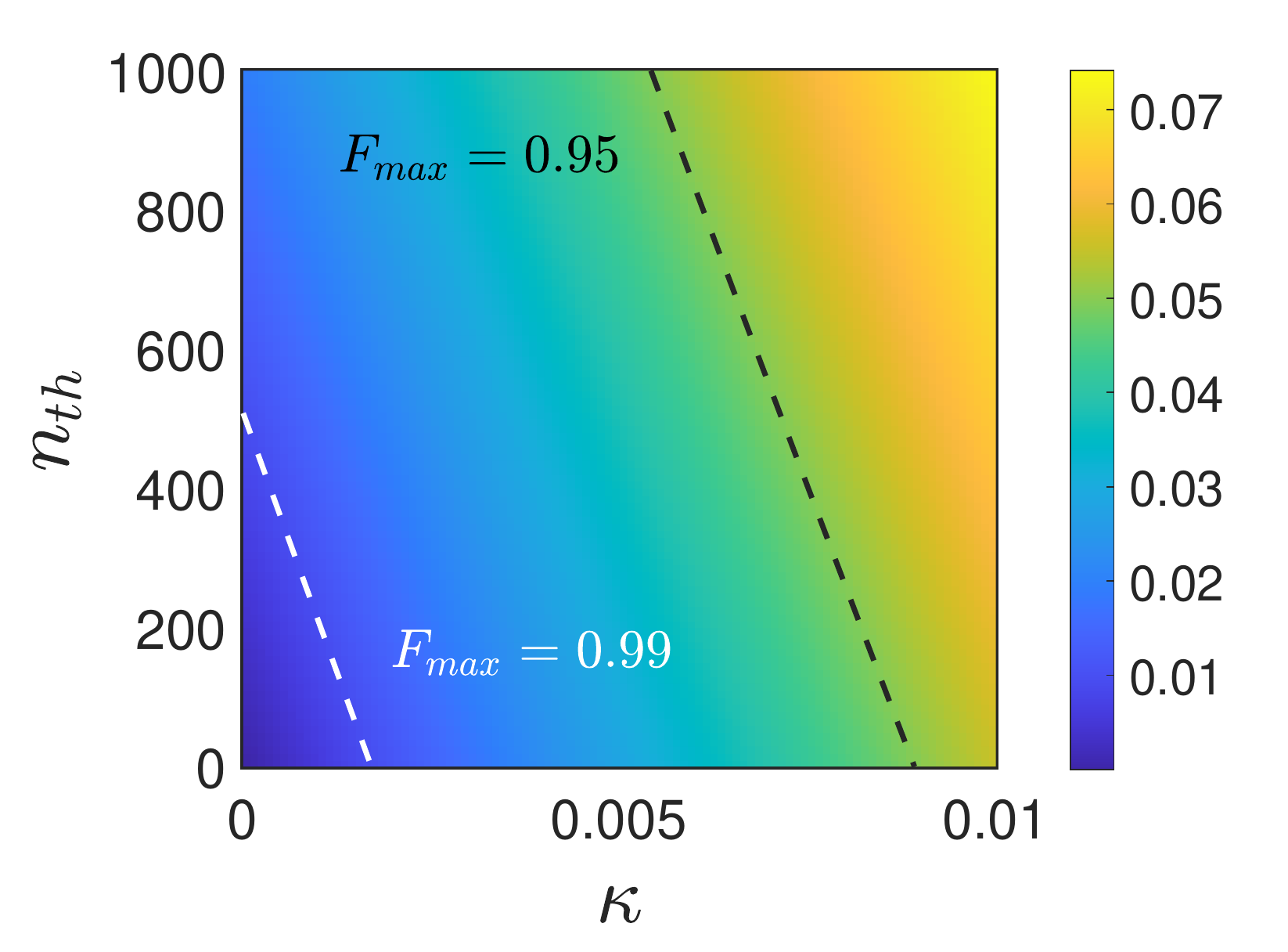}\\
  \caption{The average fidelity of controlled phase gate as the function of cavity dissipation $\kappa$ and the thermal excitation number of phonons $n_{th}$.
  The white dashed-line represents $F_{max}=0.99$ and the black dashed line represents $F_{max}=0.95$.
  Other parameters are $\Delta_c'=2 \omega_m$, $g/\omega_m=10^{-3}$, $V/\omega_m=0.046$, $\omega_A/\omega_m=0.998$,
  $\gamma/\omega_m=10^{-5}$, $\gamma_A=10^{2}\gamma$.
}\label{fig5}
\end{figure}

The optomechanical system is also affected by dissipation, except for the coherent evolution described by the Hamiltonian above.
The photon in the cavity has a dissipation ratio of $\kappa$, and a loss exists when the resonator is coupled to a heat bath with an average heat occupancy rate.
These dissipation processes were captured by Liouvillian in the form of Lindblad.
Therefore, the dynamic representation of our system can be described by the following master equation,
\begin{eqnarray}
\dot{\rho}&=&-\frac{i}{\hbar}[\rho,H]+\kappa\mathcal{D}_a\rho\\&&+\sum_{j=1}^2\Big[\gamma_{j}(n_{thj}+1)\mathcal{D}_{b_{j}}\rho+\gamma_{j}n_{thj}\mathcal{D}_{b_{j}^{\dagger}}\rho]\nonumber,
\end{eqnarray}
where $\mathcal{D}_o\rho=o{\rho}o^{\dagger}-(o^{\dagger}o{\rho}+{\rho}o^{\dagger}o)/2$ is the standard Lindblad operator.
By simulating the quantum master equation, the obtained average fidelity $F_1$ of CPFG is plotted in \fref{fig5}.
We show the evolution of the average fidelity $F$ with temperature described by $n_{thj}=n_{th}$ and cavity dissipation $\kappa$.
By keeping the temperature and dissipation within the white dashed line in the figure, the average fidelity can reach $F_{max}=0.99$.
This phenomenon is due to the fact that the enhancement of the effective coupling accelerates the phase accumulation in the CPFG realization, which well suppresses the action time of thermal noise and dissipation thus reducing its impact on the fidelity.

\subsection{Photon-phonon control under strong driving condition}
The CPFG between photon and phonons demonstrated above can be actively manipulated according to the effective Hamiltonian under the weak driving condition.
However, one of the controlled operation we need in quantum cloning is CPFG between phonon and phonon, which cannot be realized directly by employing this Hamiltonian since there is no Kerr nonlinear effect between phonons.
Fortunately, the state transfer between photons and phonons have been well studied \cite{PhysRevA.87.053818}.
High-fidelity state transfer can be achieved due to the fact that the linearized coupling coefficient dominating the state exchange is well enhanced by the average number of photons inside the cavity under the strong driving condition.
Therefore, together with photon-phonon CPFG and photon-phonon state transfer, the CPFG between phonons can be achieved.

To study the dynamics of our system under the strong driving condition, we make use of the linearized approximation by decomposing the operators into the classical and quantum components \cite{RevModPhys.86.1391}, i.e., $a\rightarrow \alpha+a$ and $b_{Aj}\rightarrow \beta_j +b_{Aj}$ (define $\alpha \equiv \langle a \rangle $ and $\beta_j \equiv \langle b_{Aj} \rangle $).
The time evolution of the annihilation operators of the system in the Heisenberg picture is then governed by
\begin{eqnarray}
\dot{a} &=& -(i \Delta_c+\frac{\kappa}{2}) a+i \sum_j G_j b_{Aj}+\sqrt{\kappa} a_{in},\nonumber\\
\dot{b}_{Aj} &=& -(i\omega_{mj}'+\frac{\gamma_{Aj}}{2})b_{Aj}+i G_j^* a-i V_j b_{j}+\sqrt{\gamma_{Aj}} b_{in}^{Aj},\nonumber \\
\dot{b}_{j}  &=& -(i \omega_{mj}+\frac{\gamma_{j}}{2})b_{j}-i V_j b_{Aj}+\sqrt{\gamma_{j}}b_{in}^{j},\label{eqlin}
\end{eqnarray}
where $G_j=2\alpha \beta_j^*g_j$, $\Delta_c=\Delta_c'-\sum_j 2g_j |\beta_j|^2$, $\omega_{mj}'=\omega_{Aj}-2g_j |\alpha|^2$.

Using the method in \sref{sec3a}, one can eliminate the auxiliary modes $b_{Aj}$ and obtain the effective Hamiltonian with linearization (see Appendix~\ref{EHSD} for details):
\begin{eqnarray}\label{hlin}
H_{lin} &=& \omega_c' a^{\dag}a +\sum_{j=1}^2 \left[\omega_{j} b_{j}^{\dag}b_{j}+(G_j'a^{\dag} b_{j}+h.c.)\right],
\end{eqnarray}
where
\begin{eqnarray}
\omega_c' &=& \Delta_c-\sum_j \frac{|G_j|^2(\omega_{mj}'-\Delta_c)}{(\omega_{mj}'-\Delta_c)^2+(\gamma_{Aj}-\kappa)^2/4},\nonumber\\
\kappa' &=& \kappa+\sum_j \frac{|G_j|^2(\gamma_{Aj}-\kappa)}{(\omega_{mj}'-\Delta_c)^2+(\gamma_{Aj}-\kappa)^2/4},\nonumber\\
\omega_j &=& \omega_{mj}+\frac{V_j^2(\omega_{mj}'-\omega_{mj})}{(\omega_{mj}'-\omega_{mj})^2+(\gamma_{Aj}-\gamma_{j})^2/4},\nonumber\\
\gamma_j' &=& \gamma_{j}-\frac{V_j^2(\gamma_{Aj}-\gamma_{j})}{(\omega_{mj}'-\omega_{mj})^2+(\gamma_{Aj}-\gamma_{j})^2/4},\nonumber\\
G_j' &\approx& \frac{ G_j V_j}{\omega_{mj}'-\omega_{mj}}.\nonumber
\end{eqnarray}

It is recognized that the $H_{lin}$ in \eref{hlin} is a typical Hamiltonian of state exchange \cite{OC.430.385}.
Driving $H_{lin}$ and using the general method of state exchange, we can obtain the transport coefficient $T$ of the state between mechanical mode and optical mode,
\begin{equation}
    T=\frac{Q(t)}{U_{swap}Q_{in}},
\end{equation}
where $Q(t)=e^{M t}Q(0)+\int_0^t d \tau e^{M (t-\tau)} Q_{in}(\tau)$, $Q(t)$ denotes the state of the system mode $[ a,b_1,b_2]^T$ at the time $t$, $Q(0)$ denotes the initial state of the system (We usually prepare the initial state of the system to the ground state), $Q_{in}$ denotes the input state of the system, $U_{swap}$ denotes the swap gate operation we need to be implemented and the matrix $M$ is given by
\begin{eqnarray}
M&=& \left[
                  \begin{array}{ccc}
                    -(i\omega_c'+\frac{\kappa'}{2}) & -iG_1' & -iG_2'\\
                    iG_1'^* & -(i\omega_1+\frac{\gamma_1'}{2}) & 0\\
                    iG_2'^* & 0 & -(i\omega_2+\frac{\gamma_2'}{2})
                  \end{array}
                \right],\nonumber\\
\end{eqnarray}
when $T=1$ indicates that a perfect state exchange between two modes is available, and $T=0$ indicates that a state exchange between two modes is not available.

\begin{figure}
  \centering
  \includegraphics[width=8.2cm]{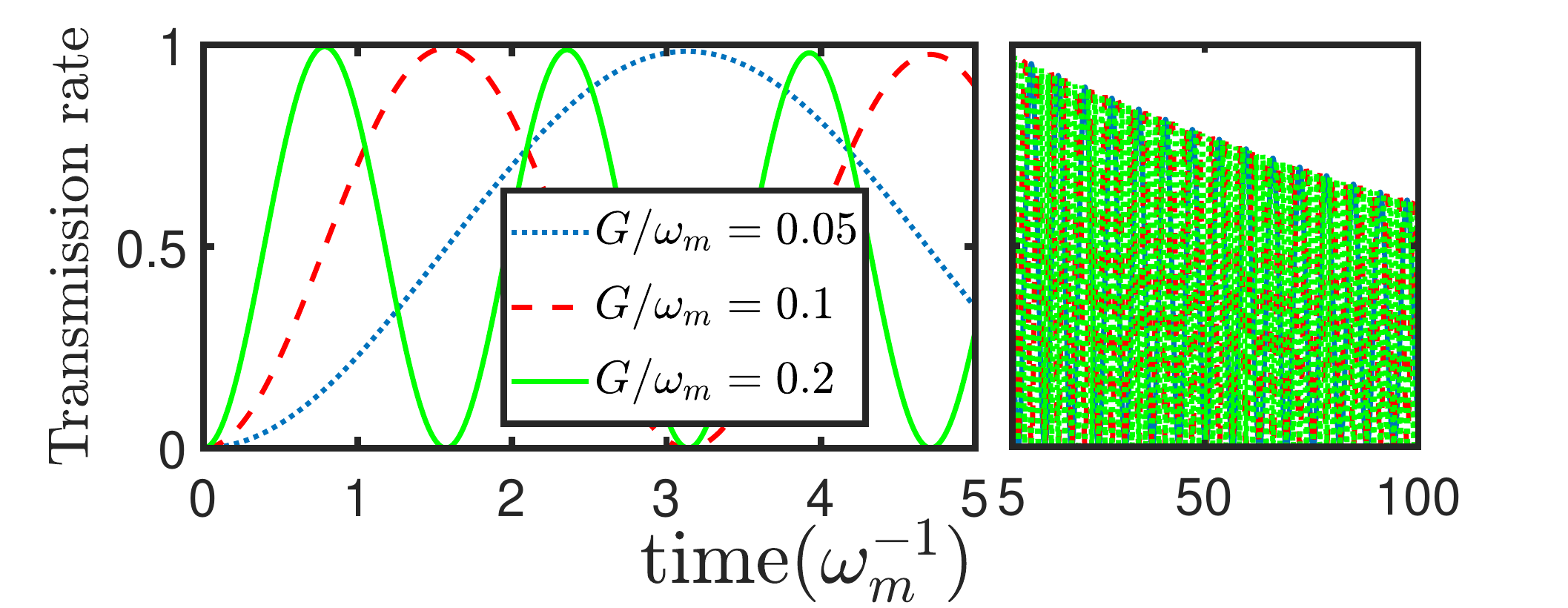}\\
  \caption{The transmission rate between optical mode $a$ and mechanical modes $b_{j}$ with different linearized coupling rates $G$.
  $\omega_c'=\omega_m$, $\kappa/\omega_m=0.1$, $n_{th}=10$.
  Other parameters are the same with \fref{fig5}.
}\label{fig6}
\end{figure}
The result of the transmission rate between optical mode $a$ and mechanical modes $b_{j}$ is explored in \fref{fig6}.
$G_j'$ is the key parameter to control the transmission speed in our scheme.
For the convenience of discussion, we set $G_j'=G$, $\omega_{mj}'=\omega_m$, and show the transmission dynamic for different coupling coefficients in the figure.
The time cost required to complete the quantum state transmission process decreases as the increase of $G/\omega_m$.
In addition, the shorter transmission time means the less dissipation impact.
Therefore, we can obtain high state transmission fidelity by enhancing the linearized coupling coefficient $G$ with enhancement of the driving strength.
Based on this state exchange between photon and phonon, combined with the CPFG between photon and phonon, a suitable quantum process design can achieve the CPFG between phonon $b_1$ and phonon $b_2$, which fulfill the requirements of quantum cloning.

\subsection{Photon-phonon quantum logic gate}

\begin{figure}
  \centering
  \includegraphics[width=8.2cm]{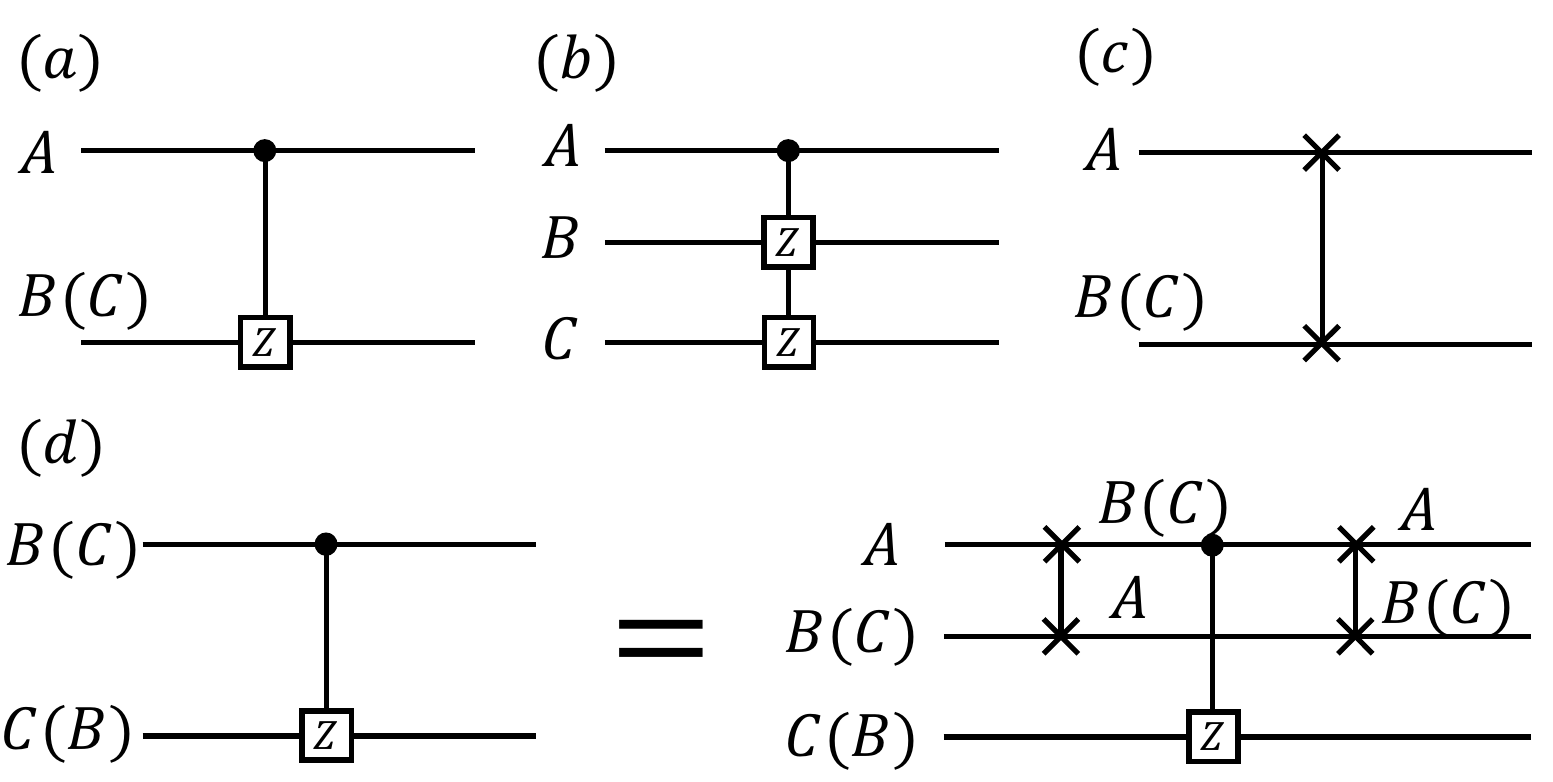}\\
  \caption{Quantum gates implemented by our model.
  (a) Photon-phonon quantum controlled gate.
  (b) Photon-phonons bilateral quantum controlled gate.
  (c) Photon-phonon quantum swap gate.
  (d) Phonon-phonon quantum controlled gate.
  We use A denotes the optical mode $a$, while B and C denote the mechanical modes $b_1$ and $b_2$, respectively.
  }\label{fig7}
\end{figure}
As shown in \fref{fig7}, we have constructed three types of photon-phonon quantum logic gates in the previous section, i.e., (a), (b) and (c), in which $A$ denotes optical mode while $B$ and $C$ denote mechanical modes $b_1$ and $b_2$, respectively.
However, to perform quantum cloning, it is not enough to have only three types of multi-qubit gates of (a), (b) and (c).
We also need a controlled gate between $b_1$ and $b_2$.
Although this controlled gate cannot be directly implemented in the previous discussion, it can be implemented through the three existing gates, as illustrated in \fref{fig7}~(d).

Next, we expect to use the phonon-photon as well as phonon-phonon quantum gates in \fref{fig7} to implement quantum state cloning, which in turn accomplishes the quantum information manipulation task we mentioned in the introduction, i.e., copying the flying bits from the solid state bits for further quantum information processing.
By examining the existing quantum cloning protocols \cite{PhysRevLett.80.4999,PhysRevLett.106.180404,QIP.14.2271, epjd.72.26,CTP.31.223,PhysRevA.54.1844,PhysRevA.62.012302,PhysRevA.65.012304,PhysRevA.67.022317,PhysRevA.99.012324,ScienceAdv.3.e1601915,PhysRevLett.126.060503}, in addition to the above-mentioned multi-bit gates, we need the following single-bit gates,
\begin{eqnarray}\label{sqgate}
U^{(1)}&=& \left(
                  \begin{array}{cc}
                    \cos(\theta_2) & \sin(\theta_2)\\
                    -\sin(\theta_2) & \cos(\theta_2) \\
                  \end{array}
                \right),\nonumber\\
U^{(2)}&=& \frac{1}{\sqrt{2}}\left(
                  \begin{array}{cc}
                   1 & 1\\
                    1 & -1 \\
                  \end{array}
                \right),\nonumber\\
H&=& \frac{1}{\sqrt{2}}\left(
                  \begin{array}{cc}
                   1 & -1\\
                   1 & 1 \\
                  \end{array}
                \right),
\end{eqnarray}
where $\theta_2$ is a given parameter (see Appendix~\ref{PQCM} for details).
Physically, single-bit logic gates for phonons and photons are implemented via the local operation of the system, which can be achieved by applying suitable voltage pulses to the system through the tip electrodes \cite{Nature.464.697}.
And it is worth noting that single-bit logic gates have been extensively studied in physical implementations and are relatively straightforward to implement with high fidelity \cite{RevModPhys.79.135}.
Thus, for the simplification of our following discussion, all the single-bit logic gates used in our model are set to be perfect.

\section{Photon-phonon quantum state cloning}\label{cloning}
To make a copy of the flying bit (photon) from the solid state (phonon) in which the information is stored for further processing, we have designed a quantum cloning scheme based on our model.
However, due to the limitation of the quantum no-cloning theorem, we do not have a way to perform perfect cloning like the classical one \cite{nature.299.802}.
In response to different needs, many quantum cloning schemes have been proposed \cite{QIP.14.2271,PhysRevLett.106.180404, epjd.72.26,ScienceAdv.3.e1601915, PhysRevA.99.012324, PhysRevLett.125.210502, PhysRevLett.126.060503}, which can be roughly divided into two categories, probabilistic cloning and deterministic cloning, where the former guarantees the fidelity of the cloned state but sacrifices the success rate, and the latter guarantees the success rate of cloning but sacrifices the fidelity of the cloned state.
Next, we will use our model to give a solid-flying bit quantum cloning scheme and consider the effect of its cloning under the premise of experimental realizability and the existence of dissipation.
We investigate separately two different categories of cloning schemes.

\subsection{Probabilistic quantum cloning machine}\label{sec4.1}
\begin{figure}
  \centering
  \includegraphics[width=8.2cm]{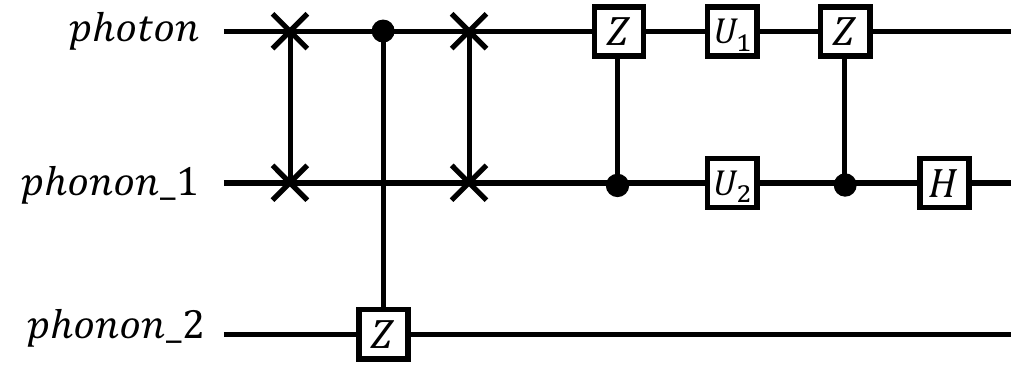}\\
  \caption{The quantum circuit of Duan-Guo PQCM for cloning the two-state set.}\label{fig8}
\end{figure}

A classical probabilistic quantum cloning machine (PQCM) for cloning the two-state set named Duan-Guo PQCM is proposed in Ref. \cite{PhysRevLett.80.4999}.
The information to be cloned is stored in phonon $b_1$, while phonon $b_2$ acts as an auxiliary bit and the information needs to be copied into the photon mode $a$.
Utilizing Duan-Guo PQCM, two-state quantum clone can be implemented (see Appendix \ref{PQCM} for details).
Using the quantum logic gates realized in our system in \fref{fig7}, we designed the quantum circuit as shown in \fref{fig8} for implementing Duan-Guo PQCM.
After taking orthogonal measurements on $b_2$, we can achieve a cloning from $b_1$ to $a$ with probability $p=1/[1+\cos(2\theta)]$, where $\theta$ is a given parameter.

\begin{figure}
  \centering
  \includegraphics[width=8.5cm]{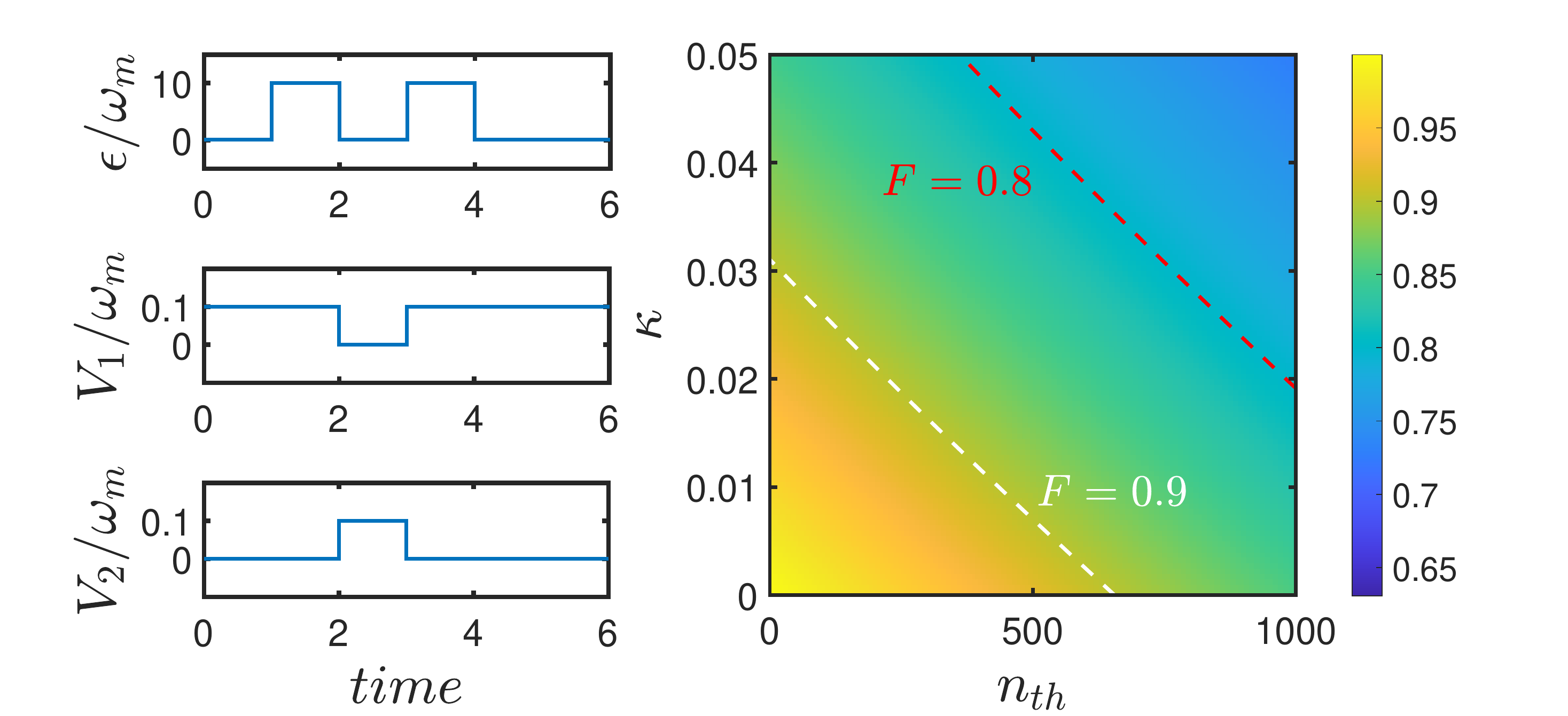}\\
  \caption{Input pulse (unit of $6.2\omega_m^{-1} $) and the corresponding output cloned state fidelity.}\label{fig9}
\end{figure}

As discussed in the previous sections, the implementation of the gates in \fref{fig8} requires different parameter conditions, hence one of the keys to implementing quantum cloning using our model is to design the corresponding adjustable parameter pulses based on the quantum circuit.
Controlled gates (weak driving) and exchange gates (strong driving) are needed in quantum cloning circuits, which requires our system be able to modulate the strength of the driving at the right time.
It can be easily achieved with current laser control devices.
Moreover, for the controlled and control bits selection, we also need to modulate the coupling strength between phonons.
In the physical realization, the coupling between two mechanical resonators can be achieved by using a piezoelectric transducer \cite{PhysRevLett.118.254301,np.9.480,PhysRevLett.117.017701}.
The mechanical coupling strength can be modulated in micro- (nano-) oscillator systems \cite{np.9.480,PhysRevLett.117.017701}.
Also can be modulated through the indirect effect of the light field \cite{np.12.27,PhysRevLett.116.103601}.
Therefore, it is enough for our model to perform this type of quantum cloning operation.
To facilitate the preparation of the pulses, we select the common period of the controlled and switched gates.
In weak driving condition, the controlled gate achieves a period of $6.2\omega_m^{-1}$.
By numerical simulation, we also find a relative period of $3.1\omega_m^{-1}$ for the exchange gate with $G/\omega_m=0.05$, i.e., a condition with driving strength $10\omega_m$, which is displayed in \fref{fig6} (see Appendix \ref{ECR} for details).
Integrated consideration, the corresponding pulsed signals with external photon driving and phonons coupling are designed as shown in \fref{fig9}.
With this pulse input, quantum cloning can be achieved.
The output fidelity of the quantum clone taking into account the environmental temperature and the system dissipation is shown in \fref{fig9}.
It is observed that the fidelity of the clone decreases with the enhancement of dissipation and environmental noise.
Thus, to achieve higher fidelity under such practical conditions, we have to reduce dissipation and environmental temperature to ensure the robustness of the system and thus improve the effect of cloning.
Note that our scheme can achieve a fidelity higher than $0.9$ under existing experimental conditions \cite{RevModPhys.86.1391} ($\kappa/\omega_m<0.03$ and $n_{th}<600$ in our scheme) without difficulty, thanks to the introduction of auxiliary modes in our model and thus to enhance the effective coupling to resist the decoherence and the environmental thermal noise.

\subsection{Deterministic quantum cloning machine}
\begin{figure*}
  \centering
  \includegraphics[width=14cm]{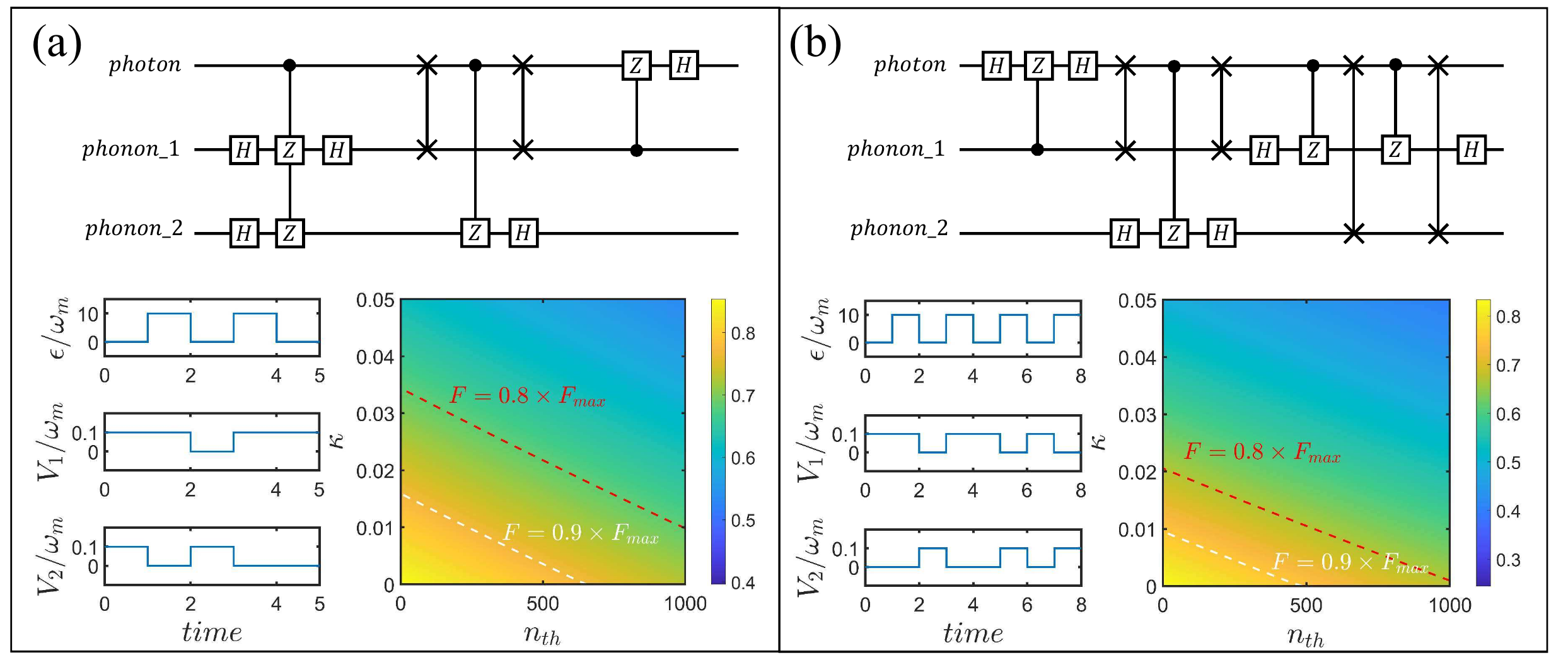}\\
  \caption{
  (a) The quantum circuit, input pulse (unit of $6.2\omega_m^{-1}$) and the corresponding output cloned state fidelity of real state quantum cloning machine.
  (b) The quantum circuit, input pulse (unit of $6.2\omega_m^{-1}$) and the corresponding output cloned state fidelity of universal quantum cloning machine.}\label{fig10}
\end{figure*}

In quantum cloning, sometimes we are more interested in the probability of success.
For this purpose, we have to sacrifice the fidelity of the clone.
A kind of deterministic quantum cloning protocols have been proposed for implementing one-to-two symmetric pair of phonon-photon quantum state copy \cite{PhysRevA.65.012304,PhysRevA.54.1844}.
Based on our model and the deterministic quantum cloning protocols, by designing specific pulses, we can implement the quantum circuits as shown in the \fref{fig10} with the consideration of dissipation.
The corresponding pulses and the clone fidelity with temperature and dissipation are displayed in \fref{fig10}.
In which, (a) and (b) denote the real-state quantum clone and the universal quantum clone, with the corresponding theoretical maximum fidelity $F^{(a)}_{max}=\sqrt{1/2+\sqrt{1/8}}$ \cite{PhysRevA.65.012304} and $F^{(b)}_{max}=5/6$ \cite{PhysRevA.54.1844}, respectively.
In a similar way to probabilistic quantum cloning in Sec.~\ref{sec4.1}, the signal to be cloned is stored in $\text{phonon}_1$ ($b_1$), and with the help of $\text{phonon}_2$ ($b_2$), we can make a copy of the information into the photon (see Appendix~\ref{DQCM} for details).
As shown in \fref{fig10}~(a), within five pulses (unit of $6.2\omega_m^{-1} $), we can implement the corresponding series of quantum gate operations and then perform quantum real-state cloning in the circuit.
Of course, to improve the copying fidelity, we still need to reduce the temperature and dissipation, and the relevant density plot is displayed in \fref{fig10}~(a).
A more general protocol to achieve universal quantum clone is shown in \fref{fig10}~(b).
The introduction of complex coefficients inevitably raises the complexity of the cloning circuit.
Thus, 8 cycles of pulses are required to carry out this clone.
More control pulses lead to more dissipation and noise, as we can see from the comparison of the two fidelity plots.
For example, if the fidelity reaches more than $0.9F_{max}$, the UQCM requires more stringent conditions.

\section{Discussion and Conclusion}\label{conclusion}
We investigate the process of cloning quantum information stored in solid qubits into flying qubits in a hybrid quadratically coupled optomechanical system.
The nonlinear interaction of the optical cavity and the mechanical oscillators cause the cavity detuning in the system.
The phase accumulation and state exchange is accelerated by increasing the values of the effective parameters $g_{j}'/\omega_{j}'$ and $G/\omega_m$ respectively.
The robustness of the system dissipation can also be improved by adjusting the parameters.
The controlled phase flip gate and bilateral controlled phase flip gate are realized in the optomechanical system under weak driving conditions, and the swap gate is realized under strong driving conditions.
Taking into account the periodic peaks of the fidelity of the control gate and the exchange gate, the optical drive pulses are set to be an integral multiple of $6.2\omega_m^{-1} $ so that they reach the peak value simultaneously at a specific time.
The quantum circuits are involved to achieve probabilistic cloning and deterministic cloning shown in \fref{fig9} and \fref{fig10}.
The probability of successful cloning by the probabilistic quantum cloning machine is sacrificed to $p=1/[1+\cos(2\theta)]$, but the fidelity can reach 1.
Considering dissipation, the fidelity after successful cloning is $0.9$.
The probability of successful deterministic cloning is 1 and the maximum fidelity can reach $F_{max}=0.854$, and the fidelity after considering dissipation is $0.9F_{max}$.

In conclusion, our hybrid optomechanical system can realize different high-fidelity quantum logic gates at specific times under different external driving strength and tunable photon-phonon coupling strength.
Using single-qubit gates and multi-qubit gates to construct quantum networks of probabilistic cloning and deterministic cloning, quantum cloning schemes that approach the theoretical optimal limit considering dissipation in designing driving pulse are realized.
By simply adjusting the pulses in \fref{fig9} and \fref{fig10}, we can perform any of the three cloning protocols in our system.
Our scheme is highly adjustable to provide great feasibility for experimental realization.
It is a new idea put forward in the direction of quantum information processing and quantum computing.
\section*{Acknowledgments}
We thank Qiu Hui-Hui, Rui-Jie Xiao and Leng Xuan for instructive discussions.
This project supported by the Natural Science Foundation of Zhejiang Province (Grant No. LY22A040005).
This project is also sponsored by the National Natural Science Foundation of China (Grant Nos. 12074206, 11704026, 11704205) and K. C. Wong Magna Fund in Ningbo University.

\onecolumn
\newpage
\appendix
\section{Approximated solution by perturbation theory}\label{ASPT}
\begin{figure}
  \centering
  \includegraphics[width=12cm]{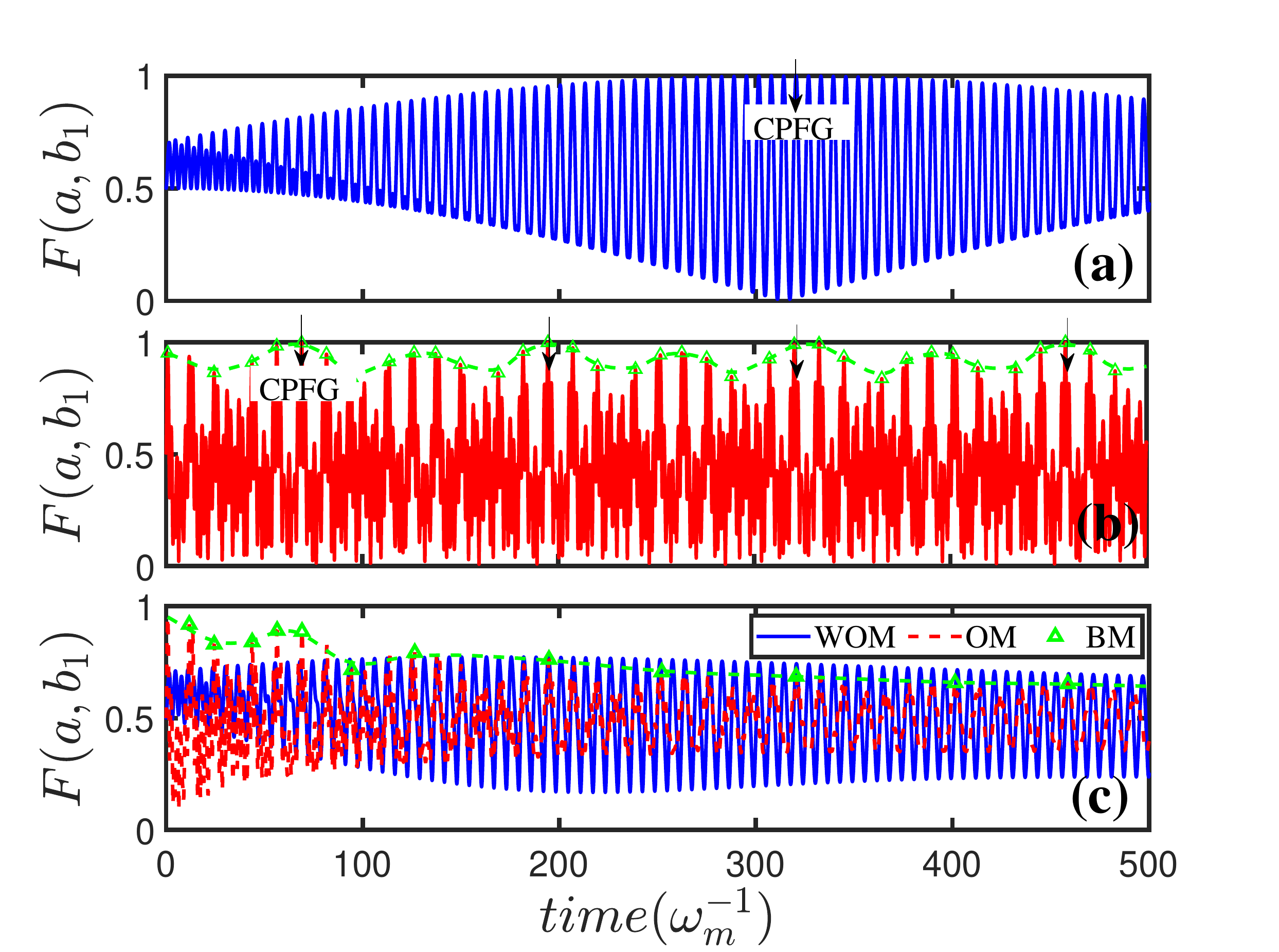}\\
  \caption{(a) The evolution of the ideal fidelity $F(a,b_1)$ of the system without outside modes (WOM) in the dissipation-free condition with $\kappa=0$.
  (b) The evolution of the ideal fidelity $F(a,b_1)$ of the system with outside modes (OM) in the dissipation-free condition with $\kappa=0$, $\omega_A/\omega_m=0.998$ and $V/\omega_m=0.03$.
  (c) The evolution of the fidelity $F(a,b_1)$ with $\kappa/\omega_m=10^{-2}$ for different model.
  BM denotes the beat frequency in dynamics.
  Other parameters are the same as in \fref{fig3}.}\label{figa1}
\end{figure}

According to perturbation theory, under condition $\{ \omega_{Aj},\omega_{mj}\}\gg V$, the solutions of $b_{Aj}$ and $b_{j}$ can be written as
\begin{eqnarray}
b_{Aj}&=&b_{Aj}^{(0)}+b_{Aj}^{(1)}+.....\nonumber\\
b_{j}&=&b_{j}^{(0)}+b_{j}^{(1)}+.....
\end{eqnarray}
where $\{b_{Aj}^{(n)},b_{j}^{(n)}\}$ denotes the solution of $n-$order of $\{b_{Aj},b_{j}\}$ with a small quantity
$\{ (\frac{V}{\omega_{Aj}})^n,(\frac{V}{\omega_{mj}})^n\}$.
By taking the form of the above solution into \eref{eqls}, we obtain the equations that $b_{Aj}^{(0)}$ and $b_{Aj}^{(1)}$ satisfy.
\begin{eqnarray}
\dot{b}_{Aj}^{(0)}&=& -(i\omega_{Aj}+\frac{\gamma_{Aj}}{2})b_{Aj}^{(0)}+2ig_j b_{Aj}^{(0)} a^{\dag}a+\sqrt{\gamma_{Aj}}b_{in}^{Aj},\\
\dot{b}_{Aj}^{(1)} &=& -(i\omega_{Aj}+\frac{\gamma_{Aj}}{2})b_{Aj}^{(1)}+2ig_j b_{Aj}^{(1)} a^{\dag}a-iV_j b_{j}^{(0)},\nonumber
\end{eqnarray}
Thus, the first-order approximate solution of $b_{Aj}$ can be obtained by
\begin{eqnarray}\label{eqbaj}
\dot{b}_{Aj}&=&\dot{b}_{Aj}^{(0)}+\dot{b}_{Aj}^{(1)}\nonumber\\
&=&-(i\omega_{Aj}+\frac{\gamma_{Aj}}{2})(b_{Aj}^{(0)}+b_{Aj}^{(1)})+2ig_j (b_{Aj}^{(0)}+b_{Aj}^{(1)}) a^{\dag}a-iV_j b_{j}^{(0)}+\sqrt{\gamma_{Aj}}b_{in}^{Aj}\nonumber\\
&=&-(i\omega_{Aj}+\frac{\gamma_{Aj}}{2})b_{Aj}+2ig_jb_{Aj} a^{\dag}a-iV_j b_{j}^{(0)}+\sqrt{\gamma_{Aj}}b_{in}^{Aj}
\end{eqnarray}
where the equation of $b_{j}^{(0)}$ is satisfied and the corresponding solution is given by
\begin{eqnarray*}
  \dot{b}_{j}^{(0)} &=& -(i \omega_{mj}+\frac{\gamma_{j}}{2})b_{j}^{(0)}+\sqrt{\gamma_{j}}b_{in}^{j},\\
\Rightarrow b_{j}^{(0)}(t)&=& b_{j}(0)e^{-(i\omega_{mj}+\gamma_{j}/2)t}+B_{in}^{j}(t).
\end{eqnarray*}
Combining the the expression of $b_{j}^{(0)}(t)$ with \eref{eqbaj} we can obtain the first order approximate solution of $b_{Aj}(t)$,
\begin{eqnarray}
b_{Aj}(t)&\approx &b_{Aj}(0)e^{-(i\omega_{Aj}+\gamma_{Aj}/2-f_N)t}+\frac{-iV_j b_{j}(t)}{i(\omega_{Aj}-\omega_{mj})+\frac{\gamma_{Aj}-\gamma_{j}}{2}-f_N}+B_{in}^{Aj}(t),\nonumber
\end{eqnarray}
where $B_{in}^{Aj}(t)=[\sqrt{\gamma_{Aj}}b_{in}^{Aj}-i V B_{in}^{j}(t)]/(i\omega_{Aj}+\gamma_{Aj}/2-f_N)$ indicates the input noise after modification.
Then, substituting the above equation into \esref{eqls} to obtain the effective
Langevin equations without $b_{Aj}$.
Thus the effective Hamiltonian can be rewritten as \eref{heff}.
To better demonstrate the advantages of introducing outside modes, we performed numerical simulations shown in \fref{figa1}, i.e., a comparison of the fidelity of CFPG $F_{a,b_1}$ with and without ($V=0$) outside modes, and for fairness, the parameters of both models were taken to be exactly the same.
As shown in \fref{figa1}(a) and (b), to demonstrate the relationship between Kerr phase and effective frequency, we compared the two models in the ideal case without dissipation.
It is observed that due to the weak single-photon coupling, the phase accumulation effect is quite weak with high frequency oscillator, and it needs lots of cycles to achieve the CPFG.
While, when the effective frequency is comparable to the Kerr coefficient, the phase accumulation becomes significant and the CPFG can be achieved in a few effective cycles (see beat frequency marked with green triangular in \fref{figa1}(b)) when introduce the outside modes.
As a result, in the presence of dissipation as well as with the same parameters ($g/\kappa=10^{-1}$), our model can achieve higher fidelity, and the corresponding conclusion is shown in \fref{figa1}(c). The fidelity of the OM model can reach as high as 0.95, while the WOM is less than 0.8.
This is consistent with the conclusions of \fref{fig2} in the main text.

\section{Effective Hamiltonian under strong driving condition}\label{EHSD}
Under the strong driving condition, using the linearized approximation, the dynamic equations for the quantum part are obtained in \esref{eqlin}.
After formal integration of the dynamic equations, we obtain
\begin{eqnarray}
b_{Aj}(t) &=& b_{Aj}(0)e^{-(i\omega_{Aj}+\frac{\gamma_{Aj}}{2})t}+\int_{0}^{t}d\tau e^{-(i\omega_{Aj}+\frac{\gamma_{Aj}}{2})(t-\tau)} \left[i G_j^*a(\tau)-iV_{j}b_{j}(\tau)+\sqrt{\gamma_{Aj}}b_{in}^{Aj}(\tau)\right], \nonumber\\
b_{j}(t) &=& b_{j}(0)e^{-(i\omega_{mj}+\frac{\gamma_{j}}{2})t}+\int_{0}^{t}d\tau e^{-(i\omega_{mj}+\frac{\gamma_{j}}{2})(t-\tau)}\left[-iV_j b_{Aj}(\tau)+\sqrt{\gamma_{j}}b_{in}^{j}(\tau)\right].
\end{eqnarray}
Under the conditon $V_j \ll \omega_{mj}$, we have $b_{j}(t) \approx e^{-(i \omega_{mj}+\gamma_j/2)} b_j(0)+\sqrt{\gamma_j} b_{in}^j$.
By taking this approximate expression of $b_{j}(t)$ into the formal integral of $b_{Aj}(t)$, the solution of $b_{Aj}(t)$ is obtained under condition $\gamma_{Aj}\gg \gamma_j$,
\begin{equation}
    b_{Aj}(t)\approx \frac{i G_j^*a}{i(\omega_{Aj}-\Delta_c)+(\gamma_{Aj}-\kappa)/2}-\frac{iV_j b_j}{i(\omega_{Aj}-\omega_{mj})+(\gamma_{Aj}-\gamma_j)/2}+\sqrt{\gamma_{Aj}}B_{in}.
\end{equation}
By inverting the above expression back into the dynamical equations \esref{eqlin}, we can eliminate $b_{Aj}$ and obtain the effective equations between $a$ and $b_j$,
\begin{eqnarray}
\dot{a}&=& -(i\omega_c'+\frac{\kappa'}{2})a-i\sum_{j=1}^2 G_j'b_j+\sqrt{\kappa'}a_{in}',\nonumber\\
\dot{b_j} &=& -(i\omega_{j}+\frac{\gamma_j'}{2})b_j -i G_j''^*a+\sqrt{\gamma_j'}b_{in}',
\end{eqnarray}
where $G_j' = \frac{i G_j V_j}{i(\omega_{mj}'-\omega_{mj})+(\gamma_{Aj}-\gamma_j)/2}$ and $G_j''^* = \frac{i G_j^* V_j}{i(\omega_{mj}'-\Delta_c)+ (\gamma_{Aj}-\kappa)/2}$.
The other effective parameters are listed below \eref{hlin}.
Under strong driving condition, we have $|\omega_{mj}'-\Delta_c| \gg 1$ and $|\omega_{mj}'-\omega_{mj}|\gg1$ due to the fact that the average number of photons $|\alpha|^2$ in the cavity is much greater than $1$ in this condition.
Namely, the contribution from the dissipative part of the expression for $G_j'$ ($G_j''^*$) in its denominator can be ignored, and in the case of choosing a suitable sideband ($\Delta_c \approx \omega_{mj}$), we can obtain that the complex conjugate of $G_j'$ is $G_j''^*$.
We can then reverse the above effective dynamical equations into effective Hamiltonian,
\begin{equation}
    H_{lin}=\omega_c' a^{\dag}a +\sum_{j=1}^2 \omega_{j} b_{j}^{\dag}b_{j}+(G_j'a^{\dag} b_{j}+h.c.),
\end{equation}

\section{Probabilistic quantum cloning machine}\label{PQCM}
Input the initial system containing randomly chosen to-be-cloned state into PQCM to perform a series of unitary operations,
\begin{eqnarray}
|\phi\rangle_{b_1}&=&\sin\theta|0\rangle_{b_1}+\cos\theta|1\rangle_{b_1},\nonumber\\
|\psi\rangle_{b_2}&=&\sin\nu_1|0\rangle_{b_2}+\cos\nu_1|1\rangle_{b_2},\nonumber\\
|\varphi\rangle_a&=&\sin\nu_2|0\rangle_a+\cos\nu_2|1\rangle_a,
\end{eqnarray}
where $\sin\nu_1\!=\!(\cos\theta_1\!-\!\sin\theta_1)\!/\!\sqrt{2}$ and $\cos\nu_1\!=\!(\cos\theta_1\!+\!\sin\theta_1)\!/\!\sqrt{2}$ with $\theta_1=1/2\arcsin[\sqrt{(1+\tan^4\theta)/2}]$, $\sin\nu_2=(\cos\theta_2-\sin\theta_2)/\sqrt{2}$ and $\cos\nu_2=(\cos\theta_2+\sin\theta_2)/\sqrt{2}$ with $\theta_2=1/2\arcsin[\sqrt{2/(1+\tan^4\theta)}+\sqrt{2/(1+\tan^{-4}\theta)}]$,
which including the be-cloned photon a, phonon $b_1$ and the probe phonon $b_2$.
Entering the initial state, and we measure the state

\begin{equation}
|\Psi\rangle_{b_1,b_2}=\sqrt{p}(\sqrt{2}\sin\theta \cos\theta|0\rangle_{b_1}-\sqrt{2}\sin2\theta_1\cos^2\theta|1\rangle_{b_1})|\psi\rangle_{b_2}+\sqrt{1-p}|1\rangle_{b_1}|\psi^{\bot} \rangle _{b_2},
\end{equation}
after input initial state passes through the phonon-phonon CPFG (see \fref{fig6} (d)), where $|\psi \rangle _{b_2}=\sin\nu_1|0\rangle_{b_2}+\cos\nu_1|1\rangle_{b_2}$ and $|\psi^{\bot} \rangle _{b_2}=\cos\nu_1|0\rangle_{b_2}-\sin\nu_1|1\rangle_{b_2}$.
If the measurement result of photon $b_2$ is $|\psi\rangle _{b_2}$, the final cloning result of operation sequence can be judged to be successful, and the probability is $\sqrt{p}$. On the contrary, if the measurement result is $|\psi^{\bot} \rangle _{b_2}$, it means that our cloning work failed with the probability is $\sqrt{1-p}$. Because of the existence of the principle of quantum no-cloning theorem \cite{nature.299.802}, the state of the phonon $b_1$ after measurement will collapse to $\sqrt{2}\sin\theta \cos\theta|0\rangle_{b_1}-\sqrt{2}\sin2\theta_1 \cos^2\theta|1\rangle_{b_1}$, and phonon $b_2$ is decoupled from the other two states to obtain the entangled state of photon a and phonon $b_1$.
The corresponding output state can be written as
\begin{eqnarray}
|\Phi\rangle_{b_1,a}&=&\sqrt{2}\sin\theta \cos\theta|0\rangle_{b_1}(\sin\nu_2|0\rangle_a+\cos\nu_2|1\rangle_a)\nonumber\\&&+\sqrt{2}\sin2\theta_1\cos^2\theta(-\sin\nu_2|0\rangle_a+\cos\nu_2|1\rangle_a).
\end{eqnarray}
Next, photon $a$ and phonon $b_1$ pass through different single qubit gates $U^{(1)}$ and $U^{(2)}$ respectively, and jointly pass through a double-bit qubit gate photon-phonon CPFG with phonon $b_1$ as the control bit, and then obtain a state
\begin{eqnarray}
|\Phi'\rangle_{b_1,a}&=&\frac{1}{\sqrt{2}}\sin\theta \cos\theta[|0\rangle_{b_1}(|0\rangle_a+|1\rangle_a)+|1\rangle_{b_1}(|0\rangle_a-|1\rangle_a)]\nonumber\\&&
+\frac{1}{\sqrt{2}}\sin2\theta_1 \cos^2\theta[|0\rangle_{b_1}(\chi_1|0\rangle_a+\chi_2|1\rangle_a)+|1\rangle_{b_1}(-\chi_1|0\rangle_a+\chi_2|1\rangle_a)].
\end{eqnarray}
where $\chi_1=\sin2\theta_2-\cos2\theta_2$, $\chi_2=\sin2\theta_2+\cos2\theta_2$. After that, the phonon $b_1$ passes through the quantum logic gate H, and $|\Phi'\rangle_{b_1,a}$ evolves into
\begin{eqnarray}
|\Phi''\rangle_{b_1,a}&=&\sin\theta \cos\theta(|01\rangle_{{b_1,a}}+|10\rangle_{{b_1,a}})+\sin^2\theta|00\rangle_{{b_1,a}}+\cos^2\theta|11\rangle_{{b_1,a}}\nonumber\\&=&|\varphi\rangle_{b_1}\otimes|\varphi\rangle_a.
\end{eqnarray}

\section{Effective coupling rate}\label{ECR}
For simplicity, we set the marks with $1$ and $2$ phonons are exactly the same, which means $\omega_{mj}=\omega_{m}$, $V_j=V$, $\beta_j=\beta$ and $G'_j=G'$.

\begin{figure}
  \centering
  \includegraphics[width=12cm]{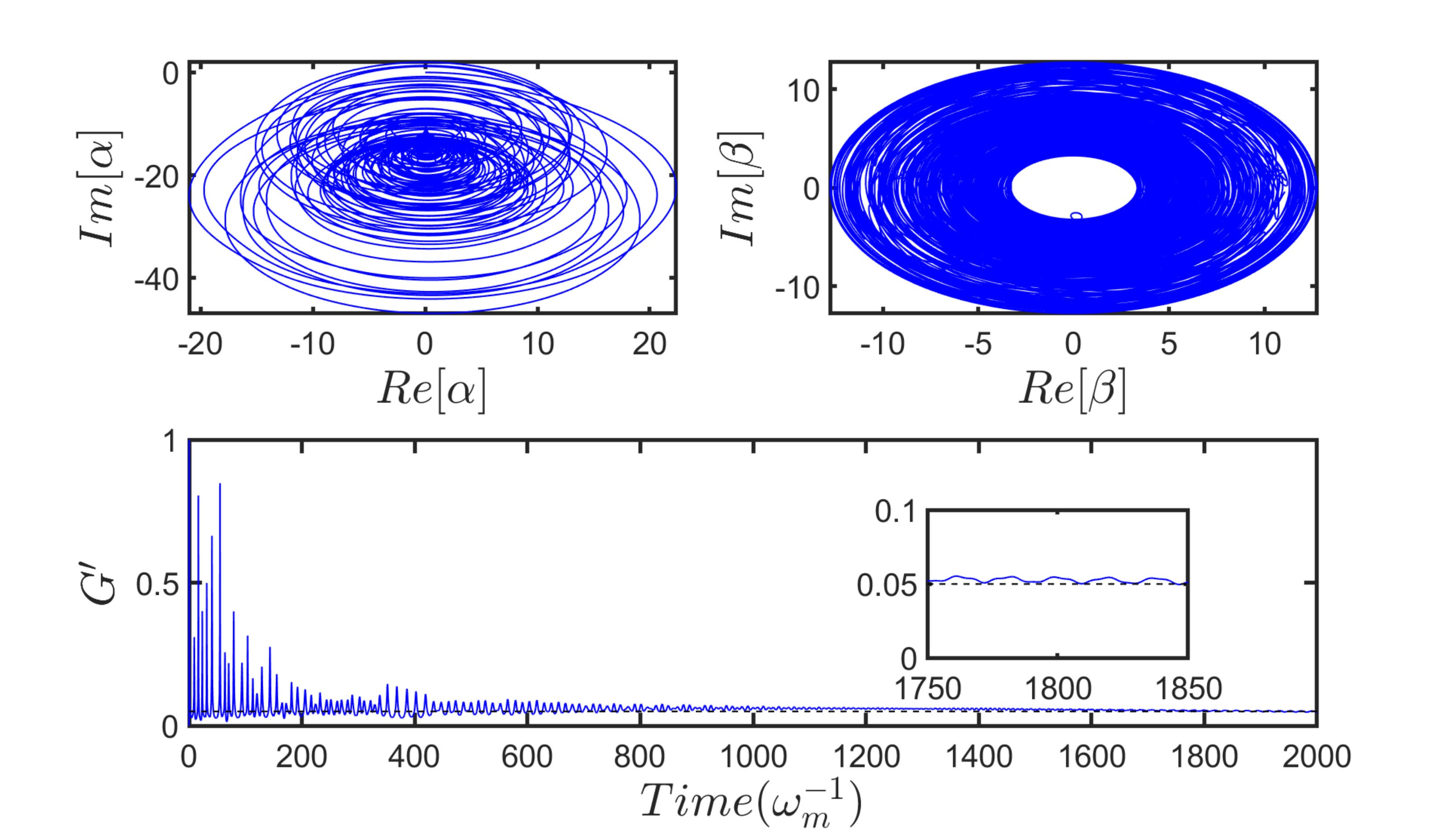}\\
  \caption{Phase of $\alpha$ and $\beta$, effective coupling rate $G'$ (unit of $\omega_m$) as a function of time.
  Other parameters are the same as those in \fref{fig2}.}\label{figa2}
\end{figure}

By numerical simulation starting with the original parameters, the average photon number $\alpha$, phonon number $\beta$ and the effective coupling rate $G'$ are displayed in \fref{figa2}.
It is observed that the average value of phonon and photon are converged to a limit loop, which means, it eventually stabilizes and the corresponding effective optomechanical coupling coefficient fluctuates around the stable value, i.e., $0.05\omega_m$ with driving strength $10 \omega_m$.
In experiments, we need to drive the system to the steady state in advance.
In addition, it is possible to accelerate this stabilization process with parameters control \cite{PhysRevLett.112.213602,PhysRevA.93.063853,PhysRevLett.116.183602}.
But, if the stability time is greater than the decoherence time, we need to prepare two hybrid optomechanical systems.
One of them using strong driving and the other using weak driving, and each of them is driven to the steady state.
Then it is enough to replace the pulse strength adjustment of the driving in the main text with a switch between this two systems.

\section{Deterministic quantum cloning machine}\label{DQCM}
In the optimal one-to-two symmetric pair of phonon-photon real state quantum clone, the cloned state is expressed as $| \psi\rangle_{b_1}=\alpha_{i} | 0\rangle+\beta_{i} | 1\rangle $.
The parameters $\alpha_{i}$ and $\beta_{i}$ in the equation are both unknown real numbers and satisfy $|\alpha_{i}|^2+|\beta_{i}|^2=1$.
Based on our model, the quantum state to be cloned is stored in the phonon $b_1$.
The auxiliary phonon $b_2$ is prepared in state $|\psi\rangle_{b_2}=\cos{\theta_1}| 0\rangle_2+\sin{\theta_1 | 1\rangle_2} $.
The photon is prepared in state $|\psi\rangle_a=\cos{\theta_2}| 0\rangle_a+\sin{\theta_2 | 1\rangle_a} $.
In which, $\theta_1$ and $\theta_2$ are adjustable known parameters and can be set according to our needs.
After performing the quantum circuit as illustrated in \fref{fig9}, the corresponding output state can be written as,
\begin{eqnarray}
| \psi \rangle_f &=& \alpha_{i} [ \cos{\theta_1}\cos{\theta_2} |000 \rangle +\cos{\theta_1}\sin{\theta_2} |100 \rangle \\
&&+\sin{\theta_1}\cos{\theta_2} |010 \rangle +\sin{\theta_1}\sin{\theta_2} |110 \rangle ]\nonumber \\
&&+\beta_i[ \cos{\theta_1}\cos{\theta_2} |111 \rangle +\cos{\theta_1}\sin{\theta_2} |011 \rangle \nonumber \\
&&+\!\sin{\theta_1}\!\cos{\theta_2} |101 \rangle \!+\!\sin{\theta_1}\!\sin{\theta_2} |001 \rangle ]_{b_1,b_2,a}.\nonumber
\end{eqnarray}

Optimal quantum state cloning can be achieved as long as we choose the appropriate parameters, where
\begin{eqnarray}
\cos{\theta_1}&=&\cos{\theta_2}=\sqrt{\frac{1}{2}+\frac{1}{\sqrt{8}}},\nonumber \\
 \sin{\theta_1} &=& \sin{\theta_2}=\sqrt{\frac{1}{2}-\frac{1}{\sqrt{8}}}.
\end{eqnarray}
In the ideal case (all the quantum gate operations are perfect ones) the corresponding optimal cloning fidelity is obtained as $F_{max}=\sqrt{1/2+\sqrt{1/8}}$ \cite{PhysRevA.65.012304}, where the fidelity is defined in the form as $F_{max}=\sqrt{\langle \Psi | \rho | \Psi \rangle}$.

For universal quantum cloning machine (UQCM), the arbitrary two-dimensional quantum state stored in the phonon $b_1$ can be expressed as $| \psi\rangle_{b_1}=\alpha_{j} | 0\rangle+\beta_{j} | 1\rangle $, where $\alpha_{j}$ and $\beta_{j}$ are complex numbers with $|\alpha_{j}|^2+|\beta_{j}|^2=1$. Phonon $b_2$ and photon $a$ are the reference state and auxiliary state, respectively, and their input state $|\Psi\rangle_{b_2,a}=s|\Phi^{+}\rangle_{b_2,a}+t|0\rangle_{b_2}|+\rangle_a$ is into the quantum circuit, where $|\Phi^{+}\rangle_{b_2,a}=\frac{1}{\sqrt{2}}(|00\rangle_{b_2,a}+|11\rangle_{b_2,a})$ is the entangled state of phonon $b_2$ and photon $a$, and $|0\rangle_{b_2}|+\rangle_a=\frac{1}{\sqrt{2}}|0\rangle_{b_2}(|0\rangle_a+|1\rangle_a)$ is the direct product state of phonon $b_2$ and photon $a$.
Optimal one-to-two symmetric pair of phonon-photon universal quantum state cloning can be achieved as long as we choose $s=t=\frac{1}{\sqrt{3}}$ \cite{PhysRevA.54.1844}.
The corresponding optimal cloning fidelity is obtained as $F_{max1}=F_{max2}=\frac{5}{6}$.

\bibliographystyle{unsrt}
\bibliography{qcom}

\end{document}